\documentclass[twocolumn,showpacs,preprintnumbers,amsmath,amssymb,superscriptaddress,noeprint]{revtex4-2}
\usepackage{graphicx}
\usepackage{amsmath}
\usepackage{amssymb}
\usepackage{orcidlink}
\usepackage{bm}
\usepackage{epstopdf}
\usepackage{diagbox}
\usepackage{xcolor}
\usepackage{float}

\begin{document}
\title{Anomalous Induced Density of Supercritical Coulomb Impurities in Graphene Under Strong Magnetic Fields}
\author{Hoang-Anh Le\,\orcidlink{0000-0002-1668-8984}}
\affiliation{Department of Physics, Korea  University, Seoul 02841, Korea}
\affiliation{Center for Quantum Nanoscience, Institute for Basic Science (IBS), Seoul 03760, Republic of Korea}
\affiliation{Ewha Womans University, Seoul 03760, Republic of Korea}
\author{S.-R. Eric Yang\,\orcidlink{0000-0003-3377-1859}}
\email{Corresponding author: eyang812@gmail.com}
\affiliation{Department of Physics, Korea  University, Seoul 02841, Korea}

\begin{abstract}
The Coulomb impurity problem of graphene, in the absence of a magnetic field, displays discrete scale invariance. Applying a magnetic field introduces a new magnetic length scale 
$\ell$ and breaks discrete scale invariance.  Moreover, a magnetic field is a singular perturbation as it turns complex energies into real energies.   Nonetheless, the Coulomb potential must be regularized with a length  $R$ at short distances  for supercritical  impurities.  
We investigate the  structure of the induced density of a filled Landau impurity band  in the supercritical regime.  
The coupling between Landau level states by the impurity potential is non-trivial and can lead to several anomalous effects. First, we find that the peak in the induced density can be located away from the center of the impurity, depending on the characteristics of the Landau impurity bands. Second, the impurity charge is  screened, despite the Landau impurity band  being filled. Third,  anticrossing impurity states lead to additional impurity cyclotron resonances.

\end{abstract}
\maketitle

\section {Introduction}

Recently, the supercritical Coulomb impurity problem has been revived \cite{ Khalilov1998,Pereira2007,Biswas2007,Shytov2007,Terekhov2008,Gamayun2009, Nishida2014,Gorbar2018,Zhu2009} in two-dimensional graphene \cite{Neto2009}.  (The problem of Coulomb impurity in three-dimensional systems was intensively investigated many years ago. For a comprehensive review, see Ref.~\cite{Reinhardt1977}.)  It has been experimentally demonstrated that single-atom vacancies in two-dimensional graphene can stably host local charge.   Using various experimental techniques, the supercritical regime can be achieved \cite{Mao2016, Wang2013}. In the absence of a magnetic field, the induced density \cite{Pereira2007, Biswas2007, Shytov2007, Terekhov2008,Nishida2014} has several interesting properties.
However, how electron-electron interactions would affect these results is not well-known.
The purpose of this paper is to investigate the induced density of the Coulomb impurity in the presence of magnetic fields.  The advantage of applying a magnetic field is that, in some cases, it reduces the effect of electron-electron interactions due to an excitation gap between the Landau levels.  We find that impurity states exhibit several unusual properties and give rise to an anomalous induced density, in addition to anticrossings that lead to new impurity cyclotron resonances.  Before we present our main findings, we provide a brief introduction of the Coulomb impurity problem both in the absence and in the presence of a magnetic field.

The continuum  model Hamiltonian of the two-dimensional  Coulomb impurity model in the absence of a magnetic field takes the following form
\begin{eqnarray}
	\mathcal{H}_0=v_F \vec{\sigma}\cdot\vec p
	-\frac{Ze^2}{\kappa r} + \Delta \sigma_z,
\end{eqnarray}
where $v_F \approx 10^6 \textrm{ m/s}$ is the Fermi velocity,  $\vec{\sigma}=(\sigma_x,\sigma_y)$ represents the Pauli spin matrices, $Ze$ is the impurity charge, $\kappa$ is the effective dielectric constant, and $\sigma_z$
is the Pauli matrix in the $z$ direction. Additionally, 
$\Delta$ represents a finite mass gap.
In the  presence of a strong Coulomb potential, to avoid pathological oscillations of wavefunctions towards the impurity origin, the impurity charge is introduced with a size of $R$.
This breaks continuous scale
symmetry into {\it discrete scale symmetry} (see Appendix \ref{App1} for an explanation of this effect).    
The coupling strength is defined as the ratio between two energy scales,
\begin{eqnarray}
	g=E_C/E_{D}=\frac{Z e^2}{\kappa \hbar v_F},
	\label{CC}
\end{eqnarray} 
where $E_C=Ze^2/\kappa R$ and $E_{D}=\hbar v_F/R.$
In the absence of a magnetic field and zero mass gap $\Delta=0$, subcritical and supercritical regimes separate at the critical coupling strength $g_c = 1/2$ \cite{Pereira2007, Shytov2007, Khalilov1998, Gamayun2009}.

Nishida \cite{Nishida2014} showed that, in the absence of a magnetic field,  discrete scale invariance in the induced density in the supercritical regime $|g|>g_c=1/2$ has the following form
\begin{eqnarray}
\rho (\vec{r})=\sum_{|J|<g}\frac{F_J(r/r_J^*)}{r^2}+ N_0\delta(\vec{r}),
\label{Peak}
\end{eqnarray}
 where $J$ is the total angular momentum and $r_J^*$ is a $J$-dependent regularization parameter. 
In the subcritical region, $|g|<1/2$, only the scale-independent first $\delta$-function term is present \cite{Pereira2007, Shytov2007, Biswas2007}, with the analytical form of $N_0$ given in Ref. \cite{Terekhov2008}.
The noteworthy feature is that the universal function $F_J(r/r_J^*)$  displays {\it log-periodic}  and discrete scale invariance \cite{Nishida2014}, characterized by
\begin{eqnarray}
	F_J(r/r_J^*) = F_J \left(e^{n  \pi / \sqrt{g^2-J^2} } r/r_J^* \right),
\end{eqnarray} 
where  $n$  is an integer.
The induced density exhibits a power-law tail, $\rho(r)\sim 1/r^2$, as $r\rightarrow \infty$.
 The role of screening \cite{Terekhov2008} in this induced density in the presence of electron-electron interactions has not been well investigated.

In the Coulomb impurity problem in magnetic fields, regardless of the value of the magnetic field, the critical dimensionless coupling remains constant, $g=g_c$ \cite{Kim2014}.
The problem  was investigated  both below \cite{Ho2000,Zhang2012,Kim2014} and above \cite{Kim2014,Sobol2016,Moldovan2017}  critical coupling $g_c$.  
The continuum  model Hamiltonian of the Coulomb impurity problem in a magnetic field reads
\begin{eqnarray}
	\mathcal{H} = v_F \vec{\sigma}\cdot \left(\vec p+\frac{e}{c}\vec{A} \right)
	-\frac{Ze^2}{\kappa r} + \Delta \sigma_z.
	\label{BHam}
\end{eqnarray}
The graphene sheet lies in the $xy$ plane, and we use a symmetric gauge with vector potential $\vec{A} = \frac{B}{2} (y, -x)$.  The first term gives rise to graphene Landau levels with magnetic length $\ell \equiv \sqrt{\hbar c / eB}$. The Landau level energy in the absence of impurity  and the gap $\Delta$  is
$\mathcal{E}_n/E_M= \text{sgn}(n)\sqrt{2|n|}$, where the magnetic energy scale is  $E_{M}=\hbar v_F/\ell$.
Discrete scale invariance is broken because of this magnetic  length scale.
The probability densities of the Landau level states in the absence of the impurity potential form rings with width $\ell$.
However, in the supercritical region, there is  another length scale, $R \ll \ell$, as the Coulomb potential must be regularized for supercritical impurity potentials \cite{Kim2014}.
There are now two relevant dimensionless parameters,
\begin{eqnarray}
g = E'_C / E_{M} = \frac{Z e^2}{\kappa \hbar v_F}  
\text{ and } R/\ell,
\end{eqnarray}
where $E'_C=Ze^2/\kappa\ell$ is the characteristic  energy scale of the Coulomb impurity. The coupling strength $g$ is the ratio between the Coulomb energy and the Landau level energy spacing. Notice that despite a magnetic field being considered here, this coupling strength is identical to the one in Eq. (\ref{CC}).
When $g$ is large, many Landau levels are {\it coupled} by the Coulomb potential.
Note that $g$ is independent of $\ell$. The other parameter  $R/\ell$  characterizes  the regularization parameter  of the Coulomb impurity.


\begin{figure}[h]
	\centering
	\includegraphics[width= \linewidth]{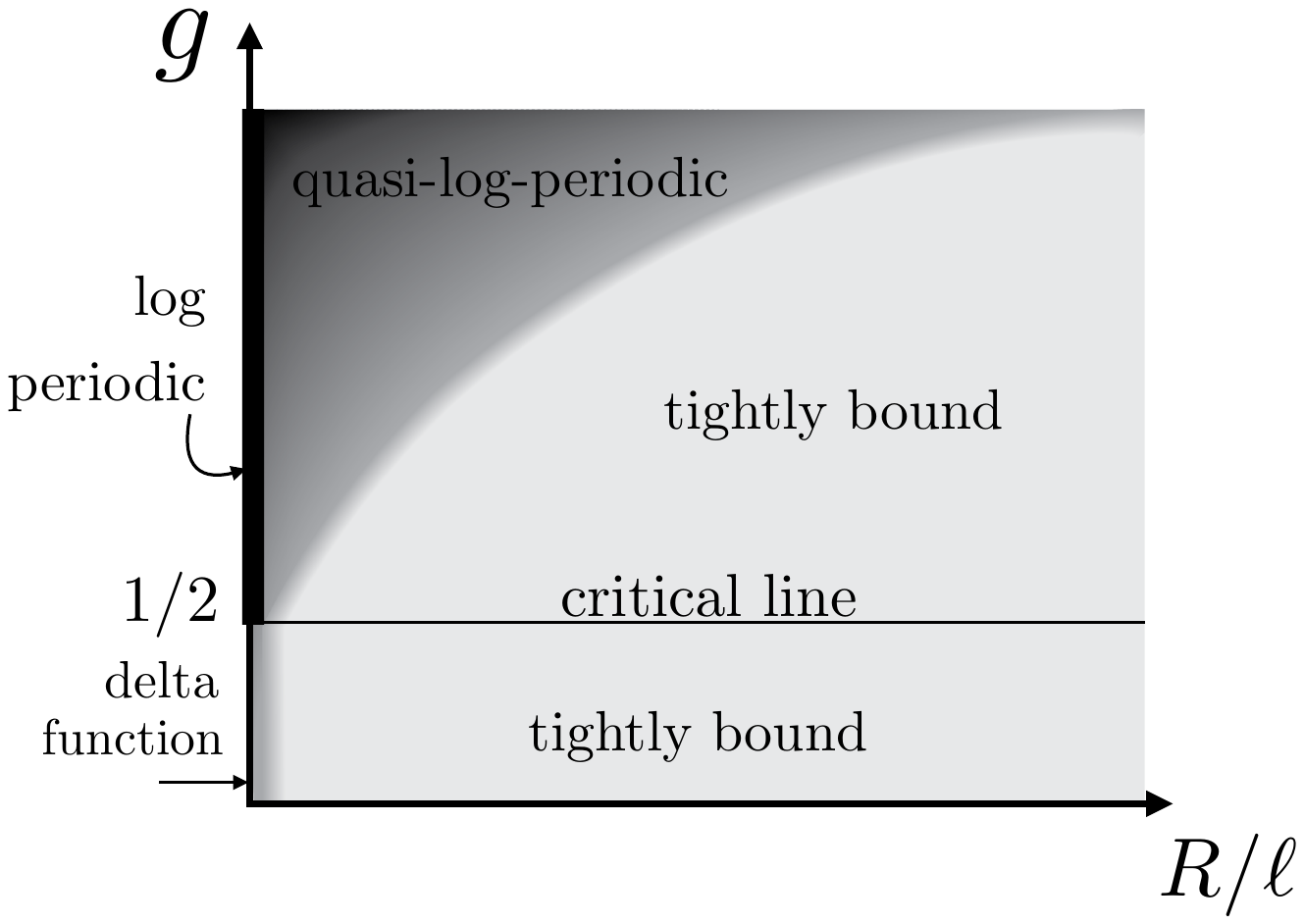}
	\caption{The wavefunctions exhibit various types of behavior: log periodic, quasi log periodic, and tightly bound.  The relevant dimensionless variables are $g$ and $R/\ell$  (see text for their definitions). There is no sharp boundary between quasi-log-periodic and tightly bound regimes. The actual shape of the ``boundary'' depends on Landau impurity band index $N$ and angular momentum $J$.   This figure is highly schematic.}
	\label{PD}
\end{figure}

In the supercritical region $g>g_c=1/2$, the following properties are found.  
 (i) No complex energy solutions (resonances) are possible in the Coulomb impurity problem in magnetic fields: the effective potential does not allow resonant states since the vector potential diverges while the Coulomb potential goes to zero in the limit $r \rightarrow \infty$~\cite{Kim2012}. 
(ii) Regardless of the size of the mass gap $\Delta$, the critical dimensionless coupling strength remains a constant $g=g_c$, unlike the case of zero magnetic field \cite{Gamayun2009}.
(iii) There can be different types of impurity bound states in a magnetic field:  quasi-log-periodic \cite{Kim2014} states for $g \gg 1$ and tightly bound states for $g\lesssim 1$, as depicted in Fig. \ref{PD}.

So far, we have briefly reviewed the basic properties of electronic wave functions of  the Coulomb impurity problem. Now, we present our main results for the induced density in the supercritical regime, particularly focusing on strong magnetic fields. We  consider  only values of the dimensionless coupling strength  $g$ where  Landau impurity bands do not overlap.   In such a case, the Landau level mixing in a filled impurity  Landau band due to many-body effects is weak~\cite{Iyengar2007}.   We investigate  how the properties of 
 induced density  in the presence of a magnetic field compare to those in its  absence.
We find that the dimensionless  induced density of such a  filled Landau impurity band  $N$ has  the following structure {\it in the supercritical region}:
\begin{equation}
\begin{aligned}
\rho_N(\vec{r}) &= \ell^2\sum_{J} |\Psi_{N,J}(\vec{r})|^2 \\
&=\ell^2\sum_{|J|\le g} |\Psi_{N,J}(\vec{r})|^2+\ell^2\sum_{|J|> g}|\Psi_{N,J}(\vec{r})|^2
\end{aligned}
\label{NewAnorm}
\end{equation}
with $\vec{r}$ being the vector position from the impurity charge.  The explicit form of $\Psi_{N, J} (\vec{r})$ will be presented below in Eq. (\ref{eigenstate}).
Here, the sum over $J$ is for all the states in the $N$th filled Landau impurity band.   
We find the following  similarities and differences in comparison to the zero-field mathematical structure of discrete scale invariance:
\begin{enumerate}
\item   
We find, as in the presence of discrete scale invariance, that states with angular momentum $J \leq g$ strongly contribute to the anomalous induced density near $r=0$.  For $N=1$ the peak in the induced density  is at $r=0$ and is most pronounced.   However, for $N=-1$ the peak in the induced density is away from the impurity center.  
The induced density displays  small oscillations for 
$r\gtrsim  \ell$, but without log-periodic oscillations for $r \gg \ell$.
\item  There is {\it} no sharp change in the peak value of $\rho_N(r)$ near the critical strength $g_c=1/2$.   The transition is smooth, but the peak value increases rapidly as $g$ exceeds $g_c$.
\item  The second term of Eq. (\ref{NewAnorm})  leads to a  unique effect  present in  a  magnetic field:  the induced density approaches  a constant value $1/(2\pi \ell^2)$ for $r > d_s$, where $d_s$ is the screening length.  (In the absence of an impurity, the density of a filled Landau level is independent of $r$ and equal to this constant value.)
\end{enumerate}  
In addition, Landau impurity band states  $\Psi_{N,J}(\vec{r})$ display anticrossings that lead to anomalous impurity cyclotro resonances.

Our paper is organized as follows. In Section \ref{section:Hamiltonian}, we explain how our numerical method is implemented using a Hamiltonian matrix.  Its eigenvalues and eigenstates that are relevant to the induced density are also explained. The properties of Landau impurity bands are explained in Section \ref{section:Landau_band}. The induced density is computed in Section \ref{section:induced_density}, and its properties are elucidated. Section \ref{section:impurity_cyclotron} explains some unusual features of impurity cyclotron resonances due to the anticrossing of Landau impurity states. Finally, discussion and a summary are given in Section \ref{section:conclusion}.


\section{Hamiltonian matrix and eigenstates}
\label{section:Hamiltonian}

We find the eigenstates and eigenvalues of the problem numerically by converting it into the diagonalization of the Hamiltonian matrix. 

We introduce the following wavefunctions to construct the basis states of the Hamiltonian:
\begin{equation}
	\psi_{n,m} (\vec{r}) = c_n \begin{pmatrix}
		-\textrm{sgn} ({n}) i \phi_{|n|-1,m} ({\Vec{r}} )      \\
		\phi_{|n|,m} ({\Vec{r}}   )   
	\end{pmatrix},
	\label{eigenvectorDirac}
\end{equation}
 where $n=\ldots ,-2,-1,0,1,2,\ldots$ and $m=0,1,2,\ldots$ are, respectively, the inter-Landau-level index and intra-Landau-level index.  These two-component states $\psi_{n,m} (\vec{r})$ are graphene Landau level states in the absence of an impurity, and their energy is given by $\mathcal{E}_n$  [see below Eq.(\ref{BHam})].  The wave function of each component $\phi_{p,q} (\vec{r})$ is defined in Appendix \ref{Twostates}.
These wave functions are defined only for $p\ge 0$ and $q\ge 0$, and are
widely used in ordinary two-dimensional gases in a magnetic field \cite{Yoshioka}. In Eq. (\ref{eigenvectorDirac}), when $|n|-1<0$ (equivalently, $n=0$), $\phi_{|n|-1,m} (\vec{r})=0$ by definition. In this case, only the second component is non zero, and the wave function is chiral.
The normalization condition of $\psi_{n,m}$ requires $c_0 = 1$ and $c_n = 1 / \sqrt{2}$ for $n \neq 0$. Note that $\text{sgn}(0) = 0$.

In constructing  the basis states  of the impurity problem in a symmetric gauge, it is useful to utilize the $z$ component of the total angular momentum:
	\begin{equation}
	J=|n|-m-1/2,
\end{equation}
as it is a good quantum number. Using $|n|=J+m+1/2$, we find that the  possible values of $J$ are half integers: $\pm 1/2, \pm 3/2, \pm 5/2, \ldots$.
(The $z$ component of the total angular momentum operator is $\hat{J}=-i\partial/\partial \theta+\sigma_z/2$, where $\theta$ is the polar angle.) Table I lists possible values of $n$ for a given value of $J$.

\begin{table}[ht!]
	\centering
	\setlength{\tabcolsep}{1em}
	{\renewcommand{\arraystretch}{2}
		\begin{tabular}{|c|cccccc}
			\hline
			& \multicolumn{6}{c}{Allowed values $n$ for a given $J$} \\ \hline
			$J \le -1/2$ & 0 & $\pm1$& $\pm2$ & $\pm 3$ & $\pm 4$ & $\cdots$\\
			$J = +1/2$ &  & $\pm1$& $\pm2$ & $\pm 3$ & $\pm 4$ &$\cdots$ \\
			$J = +3/2$ &  &  & $\pm2$ & $\pm 3$ & $\pm 4$ &$\cdots$ \\
			$\vdots$ &  &  &  & $\vdots$ &  & \\
			\hline
	\end{tabular}}
	\caption{For a given value of  $J$, the allowed values of $n$ are displayed.   For example, for $J=3/2$ the allowed values are $n=\pm2,\pm3,\ldots$, and for $J=1/2$, the possible values  are $n=\pm1,\pm2,\ldots$. Results are shown  for values of $J \leq +3/2$, with generalization to other values being obvious. \label{table_J}}
\end{table}

By relabeling $	\psi_{n,m} (\vec{r}) $  using index $J$ instead of $m$, we now introduce the {\it basis states} of the Hamiltonian matrix:
\begin{eqnarray}
	\psi'_{n,J} (\vec{r})=\psi_{n,m} (\vec{r})= \psi_{n,|n| - J -1/2} (\vec{r}).
	\label{eigenstate1}
\end{eqnarray}
The eigenstate wave function of  the Hamiltonian  with eigenenergy  $E_{N,J}$  can be expressed as a linear combination of graphene Landau level states   as follows:
\begin{eqnarray}
	\Psi_{N, J} (\vec{r}) =
	\sum_n C_n^{N,J} \psi'_{n,J}(\vec{ r}),
	\label{eigenstate}
\end{eqnarray}
 where $N$ is defined as the Landau impurity band index.
Formally, it is defined by taking the limit $g \rightarrow 0$, where an impurity state reduces to a basis state: $\Psi_{N,J} = \psi'_{n, J}$. In other words, only one term exists in Eq. (\ref{eigenstate}), and $N=n$.
The expansion coefficients $\left\{  C^{N,J}_n  \right\}$ are column eigenvectors.

For a given value of $J$,  using these basis states $	\psi'_{n,J} (\vec{r})$,  we form the total Hamiltonian matrix  in {\it the Hilbert subspace labeled by $J$}.    The relevant Hamiltonian consists of a Dirac term, the Coulomb potential, and a mass term:
\begin{equation}
\mathcal{H}_{n, n'}=	\mathcal{T}_{n,n'}+	\mathcal{V}_{n, n^\prime}  + \mathcal{D}_{n, n^\prime} .
\label{hamiltonian}
\end{equation}
Since  the graphene Landau level states are the basis, the matrix of the Dirac Hamiltonian in a magnetic field is a diagonal matrix with Landau level energies:
\begin{equation}
\mathcal{T}_{n,n'} = \text{sgn}(n)\sqrt{2|n|} \delta_{n, n'}.
\end{equation}
Here, the energy is measured in units of the magnetic energy $E_M$.  Note that employing the orthogonality in Eq. (\ref{orthogonality}), the matrix elements of the mass term can be computed as
\begin{equation}
	\mathcal{D}_{n, n'} = \frac{\Delta}{E_M} \langle{\psi_{n,m}}| \sigma_z |{\psi_{n^\prime,m^\prime}} \rangle = -\frac{\Delta}{E_M} \delta_{n , -n'}.
\end{equation}
\begin{widetext}
Matrix elements of the Coulomb potential are written as
\begin{eqnarray}
	\mathcal{V}_{n, n^\prime} &= \langle{\psi_{n,m}}| \frac{-Z e^2}{\kappa r E_M}|{\psi_{n^\prime,m^\prime}} \rangle = - 
	\langle {\psi_{n,m}}| g \frac{\ell}{r}|  {\psi_{n^\prime,m^\prime}}\rangle,
\end{eqnarray} 
which eventually can be  simplified to the following form
\begin{equation}
	\begin{aligned}
		\mathcal{V}_{n, n^\prime}= -2\pi g c_n c_{n^\prime}  \Bigg[ &\text{sgn} (n n') 2^{\alpha_1 - 1/2} A_{\vert n \vert-1, m} A_{\vert n' \vert -1, m'} I_{\beta_1 - \alpha_1/2, \beta'_1 - \alpha_1/2} \left( \alpha_1 - \frac{1}{2}, \alpha_1, \alpha_1 \right)  \\
		&+  2^{\alpha_2 - 1/2} A_{\vert n \vert, m} A_{\vert n' \vert , m'} I_{\beta_2 - \alpha_2/2, \beta'_2 - \alpha_2/2} \left( \alpha_2 - \frac{1}{2}, \alpha_2, \alpha_2 \right) \Bigg],
	\end{aligned}
\end{equation}
where $\alpha_1 = \left| J - 1/2 \right|$, $\beta_1 = \frac{2 \vert n \vert - J - 3/2}{2}$, $\beta'_1 = \frac{2 \vert n' \vert - J - 3/2}{2}$, $\alpha_2 = \left| J + 1/2 \right|$, $\beta_2 = \frac{2 \vert n \vert - J - 1/2}{2}$, and $\beta'_2 = \frac{2 \vert n' \vert - J - 1/2}{2}$. Here, $A_{n,m}$ is the normalization factor defined in Appendix \ref{Twostates} [see Eq. (\ref{Anm})]. 

To derive the above analytical form, we use the following identity of Laguerre polynomials \cite{Mavromatis1990}:
\begin{equation}
	\begin{aligned}
		I_{n,m}(\mu,\alpha,\beta) &= \int_0^\infty t^{\mu}\exp(-t) L^\alpha_m(t) L^\beta_n(t) dt \\
		&= \Gamma(\mu+1) \frac{(\beta - \mu)_n (\alpha+1)_m}{m! n!}  \;{}_3F_2\left( -m,\mu+1,\mu-\beta+1;\mu-\beta+1-n,\alpha+1
		;1\right),
	\end{aligned}
\end{equation}
\end{widetext}
where $\Gamma(x)$ is the gamma function, the Pochhammer symbol is defined as $(a)_n \equiv \Gamma(a+n) / \Gamma(a)$, and ${}_3F_2\left( a_1, a_2, a_3; b_1, b_2; 1\right)$ is the  generalized hypergeometric function. 

In the supercritical region we 
must regularize the Coulomb impurity potential by introducing  the radius of the impurity
charge $R$.  For each value of $J$, inter-Landau level numbers within $\vert n \vert \le (N_c - 1) / 2$ are included (we will call $N_c$ the Landau level cutoff).  		
The regularization parameter is related to the matrix dimension $N_c$ as follows:
\begin{equation}
	R \sim \ell\sqrt{2/N_c}.
	\label{Cut}
\end{equation}
This comes from the fact that the Landau level state with the highest index $N_c$ has this minimum length scale, namely, the distance between adjacent nodes in the wavefunction.

Some examples of the expansion coefficients $\left\{    C^{N,J}_n      \right\}$  are given in Fig. \ref{fig:cnn0jm12g09nc2501}.  Various plots of the probability densities of these eigenstates are shown in Appendix \ref{states}.

\begin{figure}[H]
	\centering
	\includegraphics[width=0.49\linewidth]{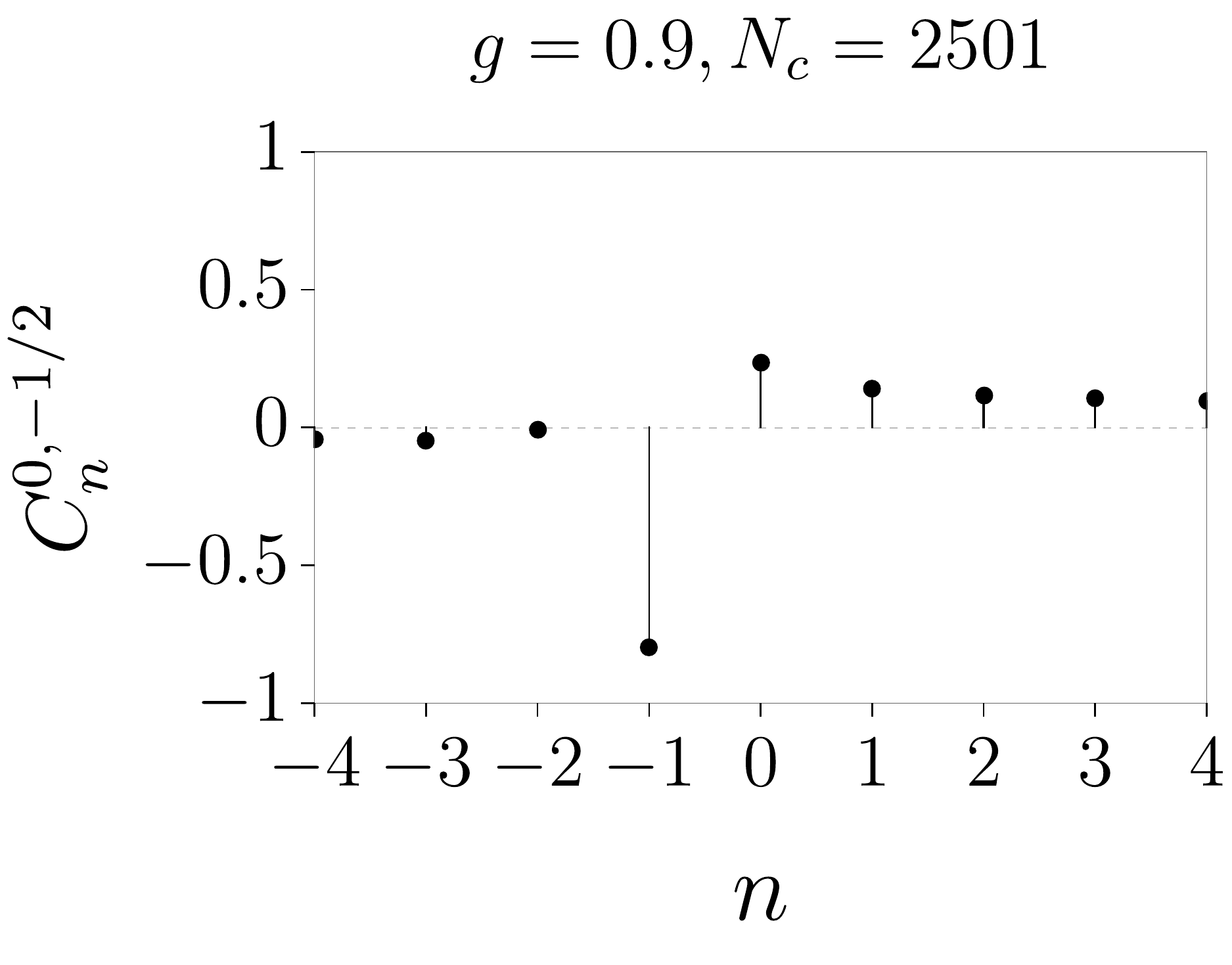}
	\includegraphics[width=0.49\linewidth]{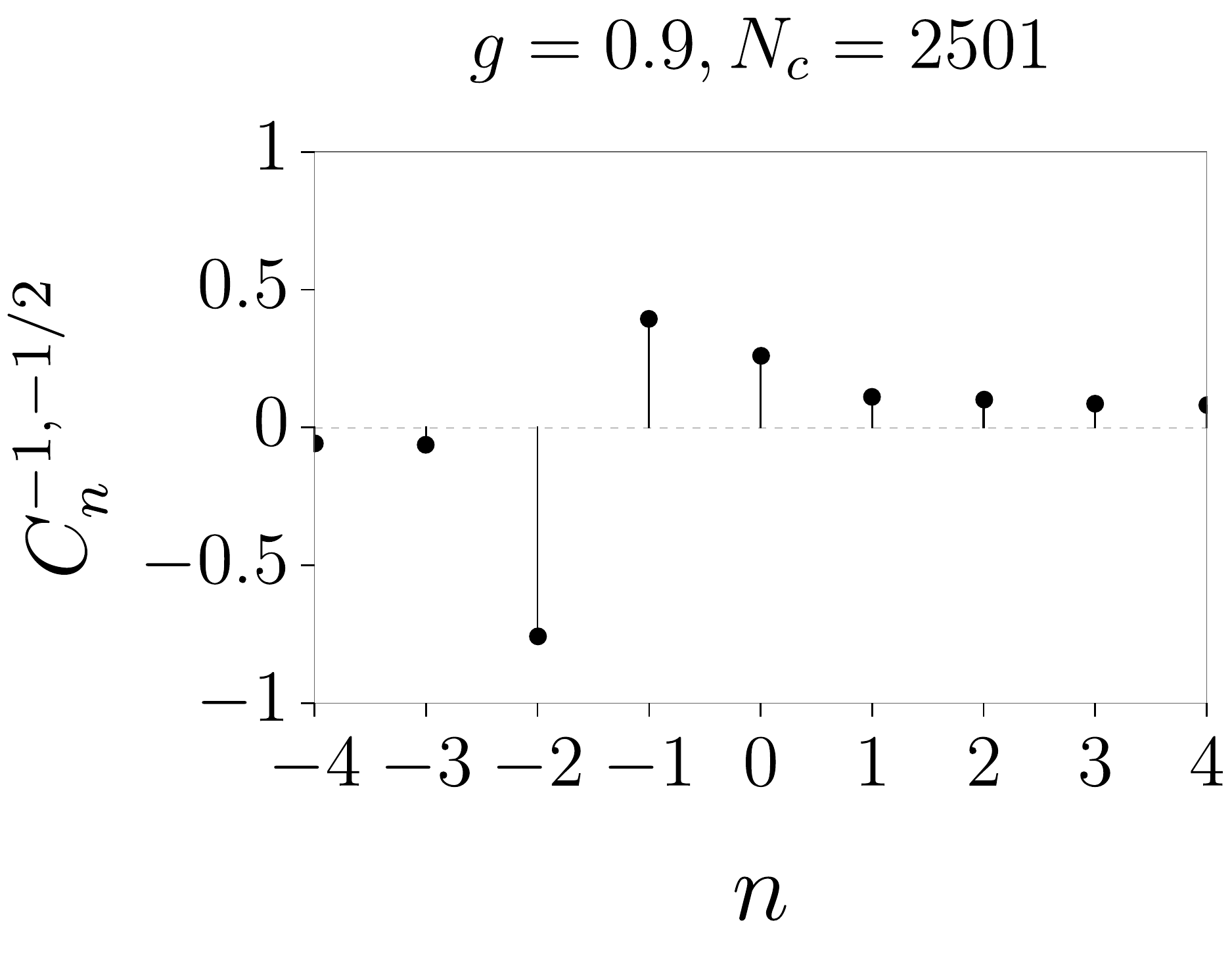}
	\caption{Expansion coefficients $C_n^{N,J} $ of Landau impurity band states as a function of $n$: $(N,J)=(0,-1/2)$   (top) and $(N,J)=(-1,-1/2)$  (bottom). With $g=0.9$, the largest contribution to impurity level $N$ comes from graphene Landau level state  with $n=N-1$.}
	\label{fig:cnn0jm12g09nc2501}
\end{figure}


\section{Landau impurity bands}
\label{section:Landau_band}

\begin{figure}[H]
	\centering
	\includegraphics[width=\linewidth]{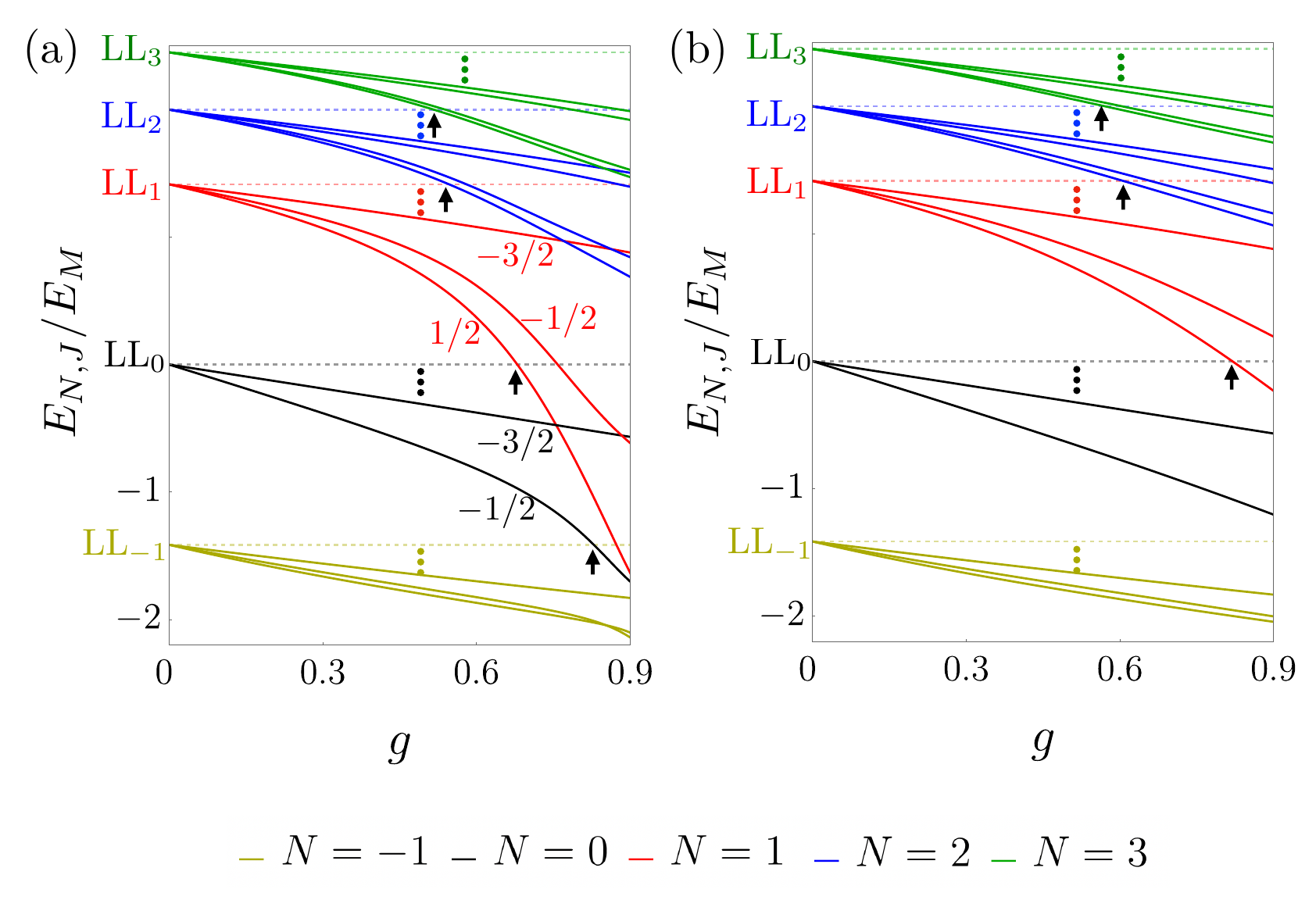}
	\caption{ 
 Landau  impurity band  states with (a) $N_c=2501$ and (b) $N_c=201$, which correspond to two different values of regularized parameter $R$. $\text{LL}_0$ stands for the  Landau  impurity band originating from the $n=0$  unperturbed Landau level described by Eq. (\ref{eigenvectorDirac}).  $\text{LL}_1$,  $\text{LL}_2$, and $\text{LL}_3$ are similarly defined.
Small numbers written next to several lines with the same color correspond to their angular momentum $J$. For each Landau level, many more energy levels are not shown, indicated by $\vdots$.    In certain cases, there exist values of $g$ where Landau impurity bands do not overlap; these values of $g$ are below the critical values indicated by the arrows.
}
	\label{oscn2}
\end{figure}

Plots of eigenvalues of Hamiltonian $\mathcal{H}$ as a function of coupling strength $g$ are presented in Fig. \ref{oscn2}. Figure \ref{oscn2}(a) corresponds to a smaller value of $R/\ell$ compared to Fig. \ref{oscn2}(b). The following points can be observed from the plot.   Firstly, an impurity splits Landau level degeneracy. This splitting, measured in units of the magnetic energy $E_M$, increases as coupling strength $g$ increases. The Landau levels $n=0$ and $n=1$ are mostly affected.
The small magnetic field $B$ limit is approached with $N_c\rightarrow \infty$, i.e., $R/\ell\rightarrow 0$ [see Eq. (\ref{Cut})]. (Our numerical approach is not suited for investigating this limit because it requires a prohibitively large value of $N_c$, meaning a prohibitively large Hamiltonian matrix.)
Secondly, there are values of $g$ where Landau impurity bands do not overlap, and we will focus on these ranges of $g$, specifically to the left of the black arrows.
Finally, in this range of $g$, the $n=0$ and $n=1$ Landau levels are strongly affected by the change in $R/\ell$. 
In addition, the Landau level splitting is {\it smaller} for a smaller value of $R/\ell$.


\section{Induced density of a filled Landau impurity band}
\label{section:induced_density}


Suppose that states of a Landau impurity band are filled, and they do not overlap in energy with other Landau impurity band states.  (There are values of $g$ for which Landau impurity bands do not overlap, positioned to the left of the black arrows in  Fig. \ref{oscn2}.)  
In such cases, mixing of Landau impurity band states  with other band states  due to many-body effects is weak, as demonstrated in Refs.~\cite{Iyengar2007,MacDonald1993}.

\subsection{Zero mass gap $\Delta=0$}

\begin{figure}[H]
	\centering
	\includegraphics[width=0.7\linewidth]{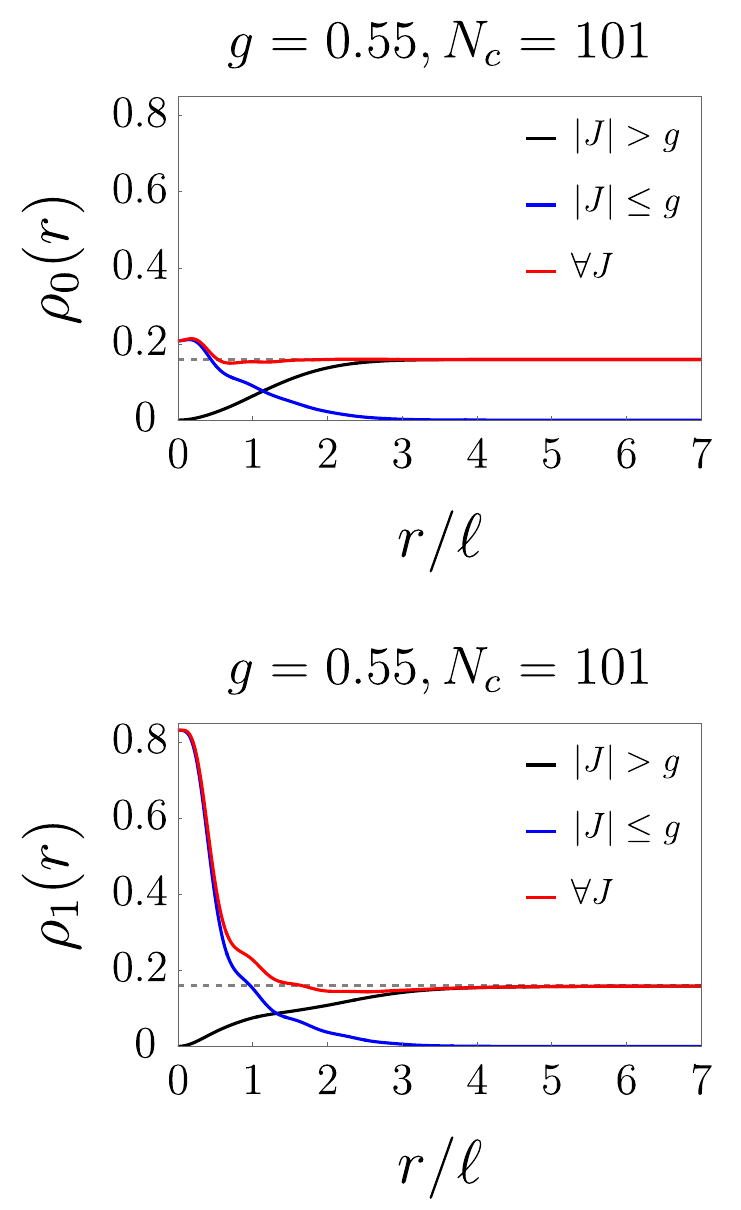}
	\caption{
 The induced charge densities at $g=0.55$ are represented by red lines for the impurity bands (top) $N=0$ and (bottom) $N = 1$.
All charge densities are computed with $N_c = 101$. Blue and black lines correspond the first and second terms of the charge density in Eq. (\ref{NewAnorm}), respectively. The $y$ axes of the two plots have a similar scale.}
	\label{fig:chargerho1g0A}
\end{figure}

\begin{figure}[H]
	\centering
	\includegraphics[width=0.7\linewidth]{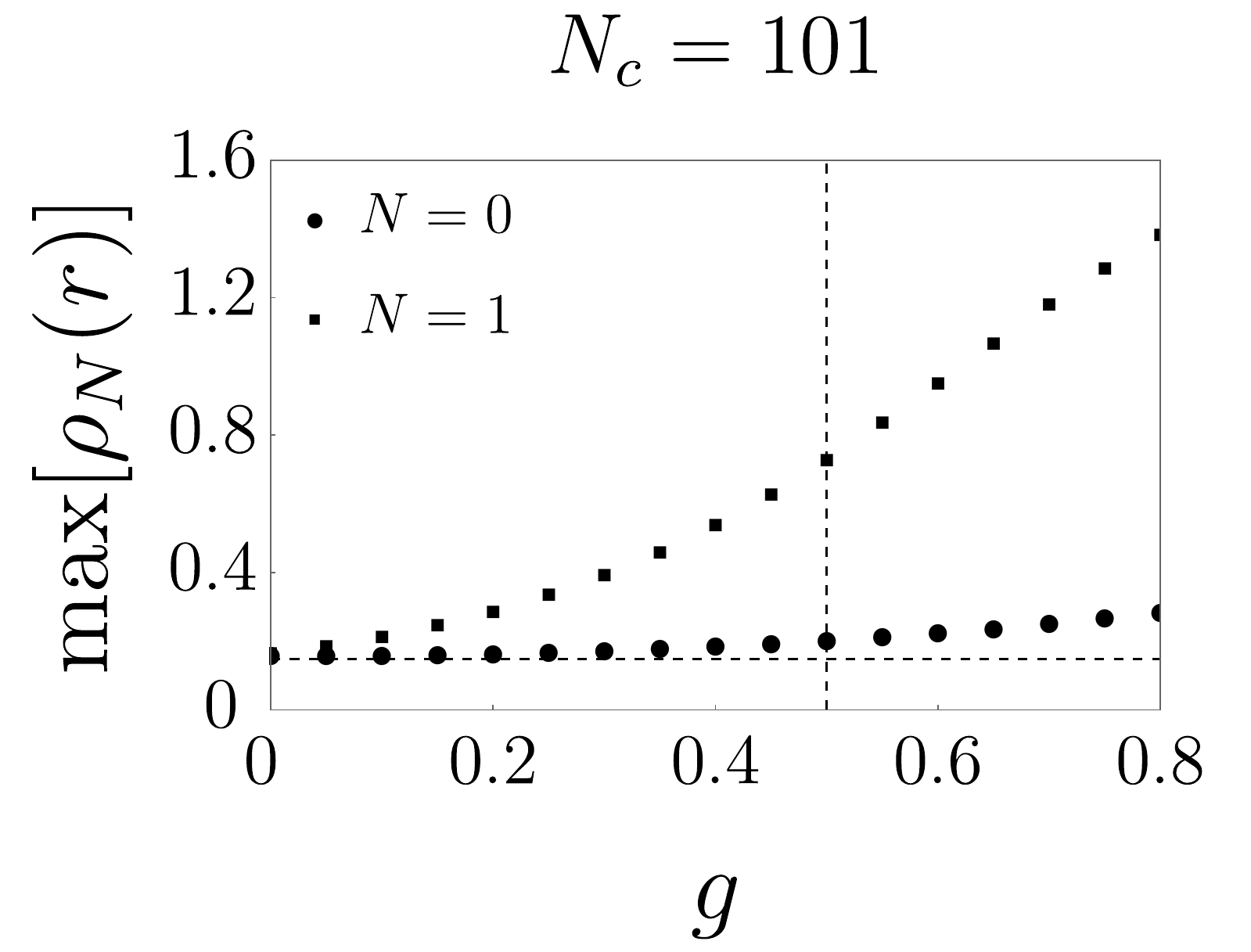}
	\caption{The peak value of $\rho_N(r)$ is computed as a function of $g$ with $N_c = 101$. The vertical dashed line indicates critical value $g_c = 0.5$. The horizontal dashed line is the constant value $1/(2\pi \ell^2)$.}
	\label{fig:maxrhog}
\end{figure}

We first investigate the massless case with $\Delta = 0$. The  behavior of the induced density of a Landau impurity band  can be  rather different from that of zero magnetic field because discrete scale invariance is not present.  We have computed the induced density in  the supercritical regime $g=0.55$ for the values of $N=0$ and $N=1$, as shown in Fig. \ref{fig:chargerho1g0A}.  We find that no $\delta$-function exists at $r=0$, but   a sharp narrowing of the induced density near the location of the impurity  is present.   
This phenomenon is a precursor of the ``fall to the center" of the electron bound to the impurity charge. 
Moreover, the position of the peak value of the induced density depends on the Landau impurity band index $N$.  For $N=0$ and $N=1 $ the peak is near  $r=0$, as shown in Fig. \ref{fig:chargerho1g0A}.
The red curves in Fig. \ref{fig:chargerho1g0A} represent $\rho_0(r)$ and $\rho_1 (r)$, while the blue and black curves represent the first and second terms of the induced density given by Eq. (\ref{NewAnorm}).  Note impurity band states with $|J|\neq 1/2$ do not contribute to the induced density at $r=0$. 
The peak value of $\rho_1 (r)$ is much {\it larger} than that of $\rho_0 (r)$.  
This is because {\it both}  Landau band impurity  states with $J=\pm 1/2$ channels, $\Psi_{1,-1/2}(\vec{r})$ and $\Psi_{1,1/2}(\vec{r})$,  contribute to it; however, for $\rho_0 (r)$,  only the state with $J = -1/2$, $\Psi_{0,-1/2}(\vec{r})$, does. 
For large $r \gg \ell$  the induced density of a filled Landau impurity band  is $1/(2\pi\ell^2)$.  This corresponds precisely to the value of a filled graphene Landau level~\cite{MacDonald1993}. There is {\it} no sharp change in the induced density as  a function of $g$ near $g_c=1/2$.  However,  the peak value increases rapidly as $g$ exceeds $g_c$. These properties of the induced density's peak are illustrated in Fig. \ref{fig:maxrhog}.

\begin{figure}[t]
	\centering
	\includegraphics[width=0.7\linewidth]{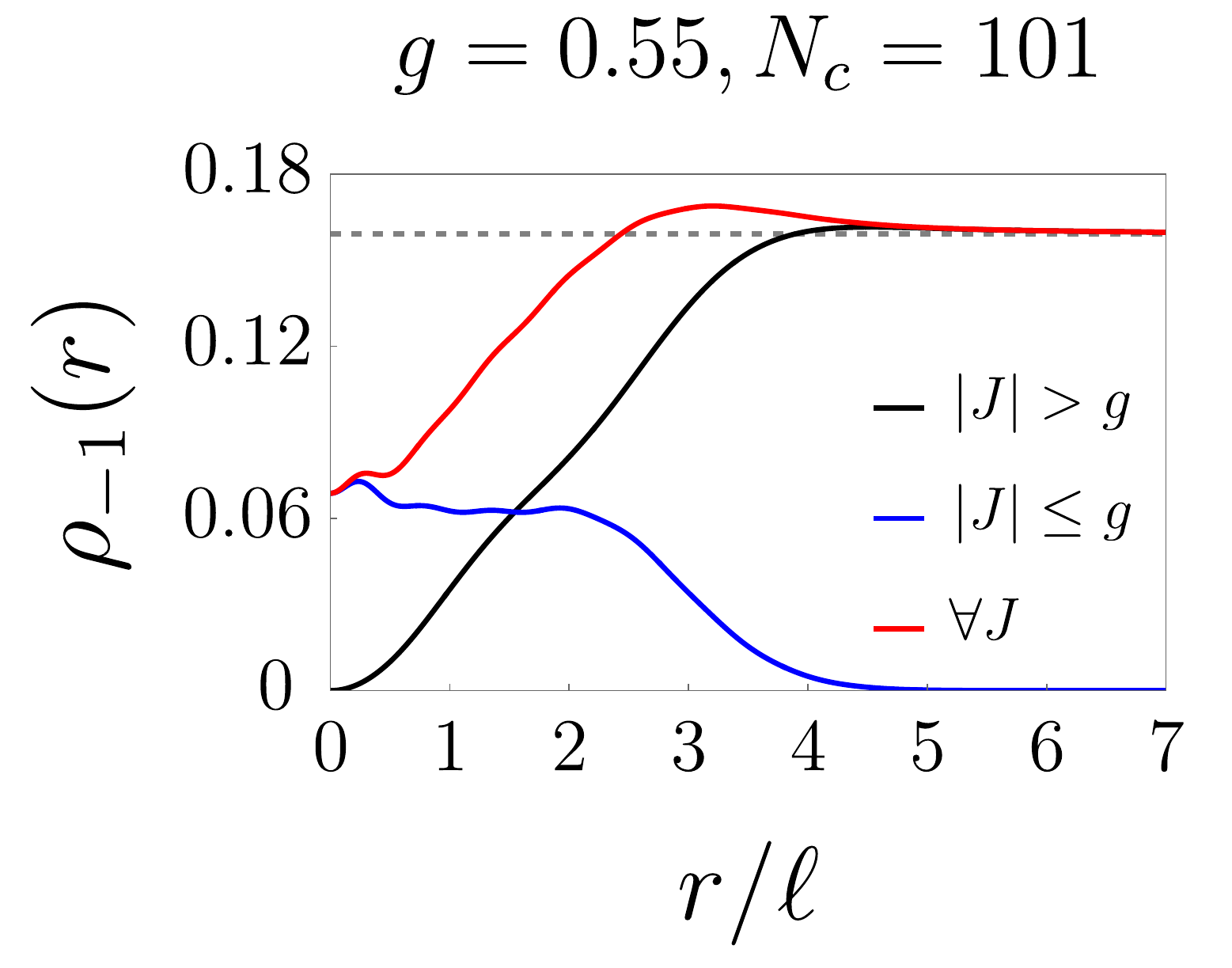}
	\caption{The induced density $\rho_{-1}(r)$ of impurity band $N=-1$ at $g=0.55$ is shown by the red line.}
	\label{n=-1} 
\end{figure}

\begin{figure}[h!]
\centering
\includegraphics[width=0.7\linewidth]{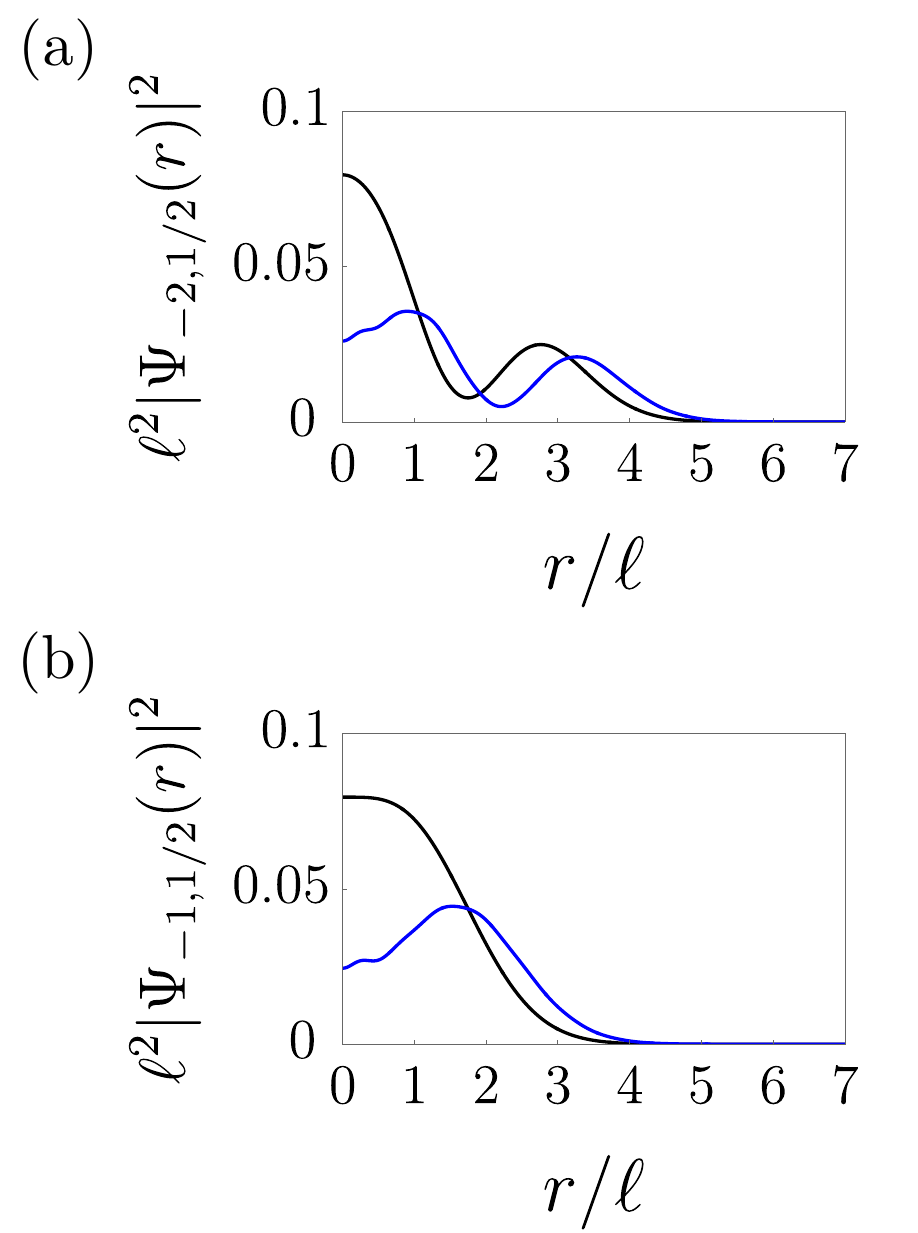}
\caption{
(a) The probability density $|\Psi_{-2,1/2}(\vec{r})|^2$ is plotted at $g=0$ (black line) and $0.55$ (blue line). (b) The same as (a) for $|\Psi_{-1,1/2}(\vec{r})|^2$. Note that at $g=0$, the impurity state $\Psi_{N,J}(\vec{r})$ is reduced to the graphene Landau level state  $\psi'_{n, J}(\vec{r})$ with $N=n$.}
\label{n=-1sec} 
\end{figure}

For some other values of $N$, the peak is located {\it away} from $r=0$. This effect becomes stronger for larger values of $g$. An example with impurity band $N=-1$ is illustrated in Fig. \ref{n=-1}: The induced density is somewhat depleted near the center but accumulates near $r\sim 2\ell$.  We can explain this effect by qualitatively analyzing the contributions of angular momentum channels to the induced density. We observe that the peak in the induced density originates from $\Psi_{-1,1/2}(\vec{r})$. According to our numerical results, this impurity state may be approximated as 
\begin{equation}
	\Psi_{-1,1/2}(\vec{r}) \approx C^{-1, 1/2}_{-1}\psi'_{-1,1/2}(\vec{r}) + C^{-1, 1/2}_{-2}\psi'_{-2,1/2}(\vec{r}),
\end{equation}
where $\psi'_{n,J}(\vec{r})$ are the basis states given in Eqs. (\ref{eigenvectorDirac}) and (\ref{eigenstate1}). These basis states are visualized by black lines in Figs. \ref{n=-1sec}(a) and \ref{n=-1sec}(b): $\psi'_{-2,1/2}(\vec{r})$ is peaked near $r\sim 2.5\ell$, while $\psi'_{-1,1/2}(\vec{r})$ is peaked at $r=0$. [Equation (\ref{posit}) provides information about the location of the wavefunctions.]  Hence, the combination of these two states causes the impurity state $\Psi_{-1,1/2}(\vec{r})$ to peak at $r \sim 2l$. The mixing between Landau levels induced by the impurity potential thus pushes these states $\psi'_{-2,1/2}(\vec{r})$ and $\psi'_{-1,1/2}(\vec{r})$  outward from the impurity center.

\begin{figure}[h!]
	\centering
	\includegraphics[width=0.7\linewidth]{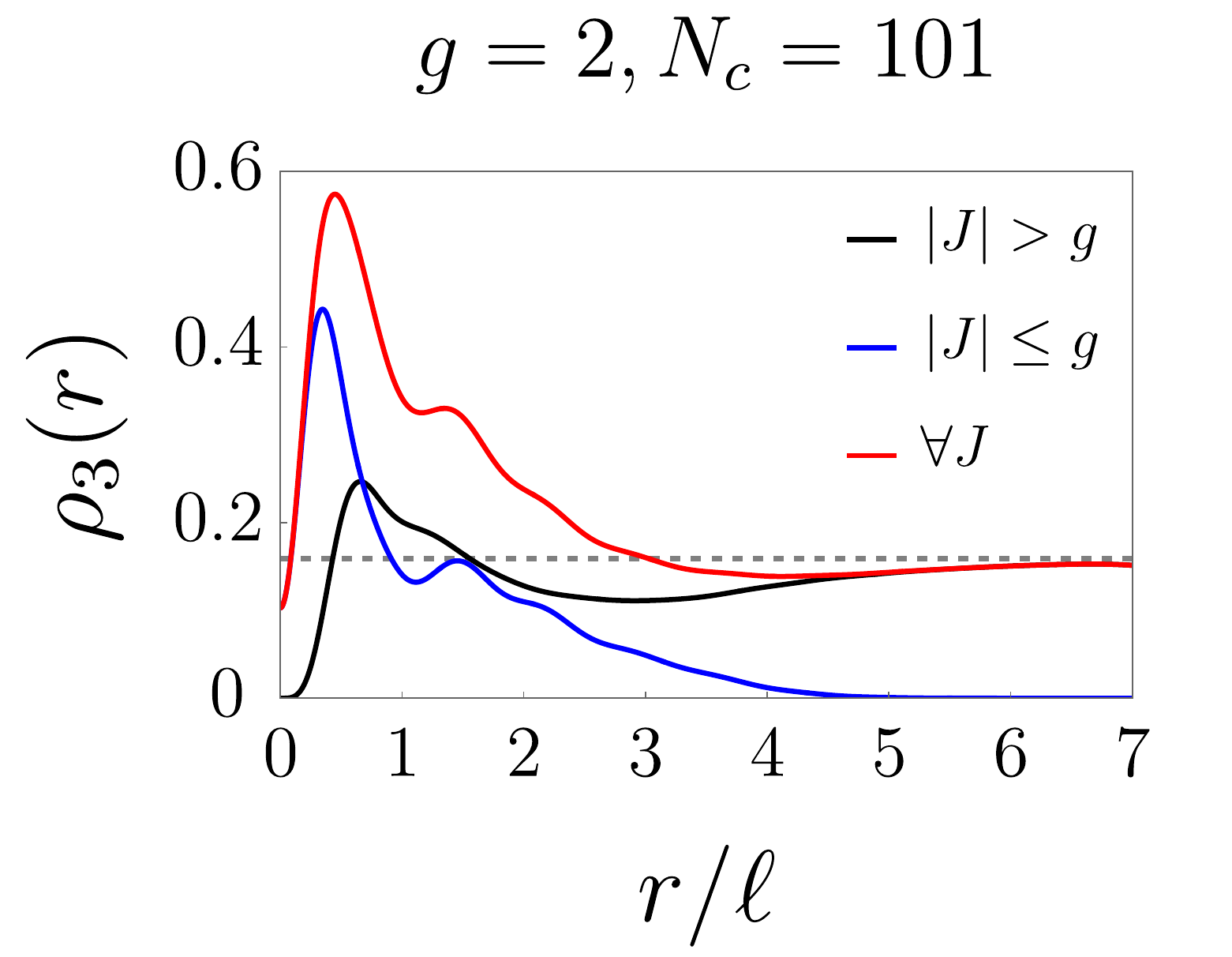}
	\caption{Induced density $\rho_3 (r)$ for impurity band $N=3$ at coupling strength $g=2$.}
	\label{fig:chargerho3g2nc101}
\end{figure}

Another noteworthy feature is that for an impurity band with a larger $|N|$ and stronger coupling strength $g$, the importance of the first term of Eq. (\ref{NewAnorm}) becomes clearer. The slope of the induced density near $r=0$ is accurately computed only when all the states $\Psi_{N, J} (\vec{r})$ with $J \leq g$ are included, as shown in Fig. \ref{fig:chargerho3g2nc101} for $N=3$ and $g=2$.

\subsection{Finite mass gap $\Delta\neq 0$}

\begin{figure}[h!]
	\centering
		\includegraphics[width=0.7\linewidth]{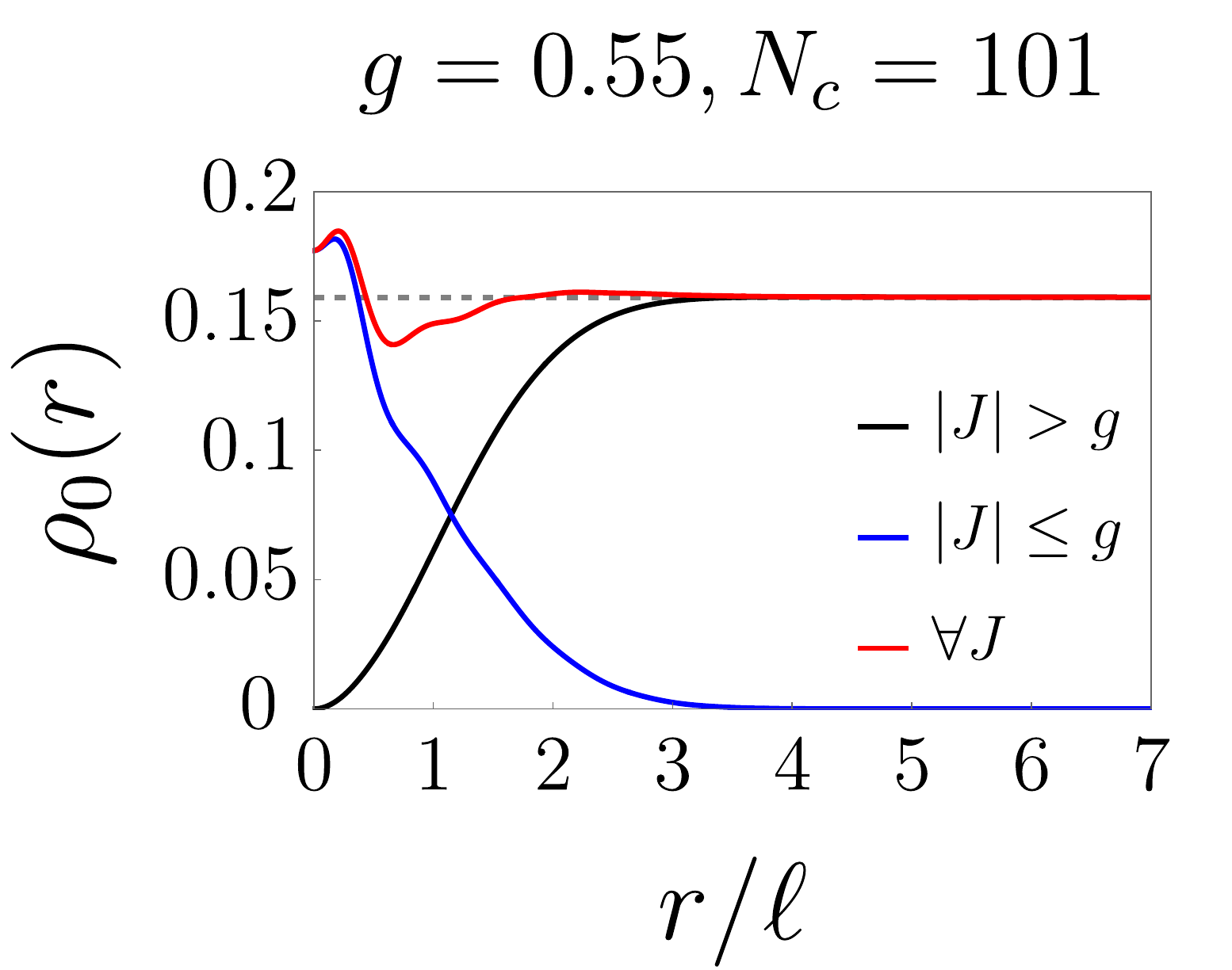}
		\includegraphics[width=0.7\linewidth]{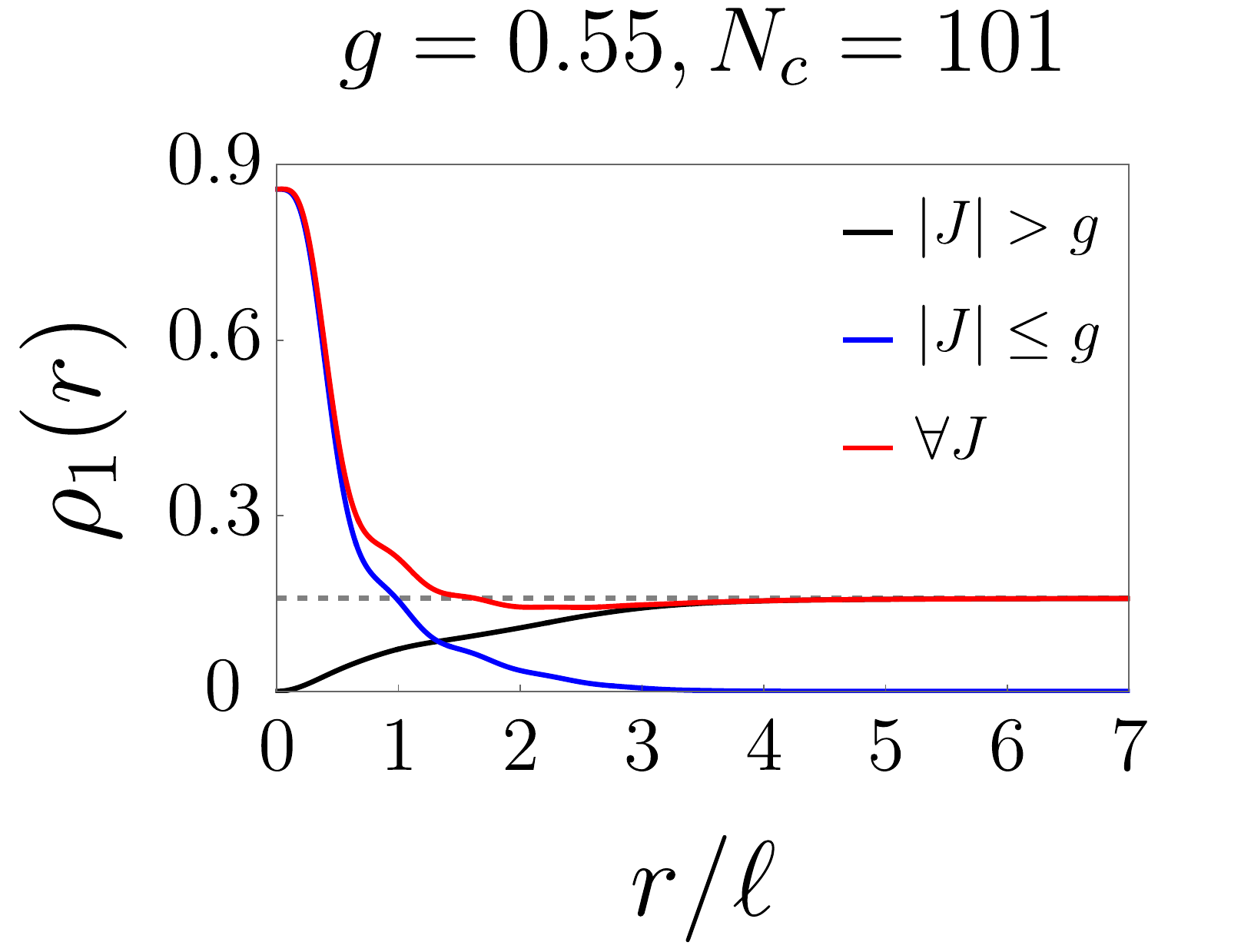}	
		\caption{Induced densities of impurity bands $N=0$ and $N = 11$ with a finite mass gap $\Delta=0.1E_M$. The coupling strength is $g = 0.55$, and matrix dimension $N_c = 101$.}
	\label{DElta}
\end{figure}

The induced densities display the same qualitative behaviors when the gap value is finite:
states with channels $|J|\le  g$ contribute
to  the peak, while other terms cause 
the corresponding induced  density  to approach a  constant 
value of $1/(2\pi \ell^2)$ at large distances.
Figure \ref{DElta} displays induced densities for $N=0$ and $N=1$ for $\Delta=0.1E_M$.

\subsection{Screening}

\begin{figure}[H]
	\centering
	\includegraphics[width=0.7\linewidth]{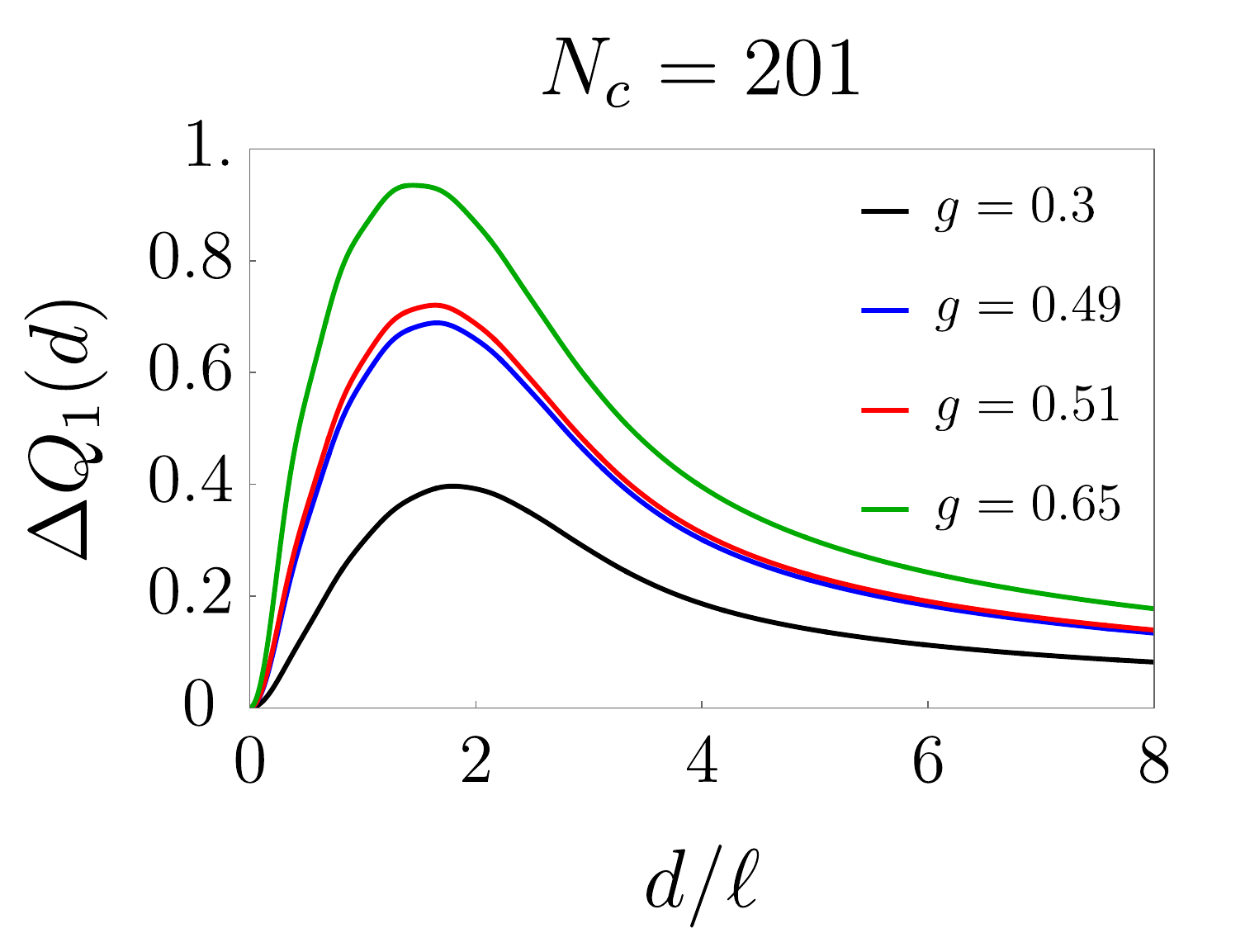}
	\caption{$\Delta Q_1 (d)$ for several values of coupling strength $g$, computed with $N_c = 201$.   Included $J$ values  are  $1/2,-1/2,-3/2,\ldots, -139/2$.   To accurately calculate $\Delta Q_1 (d)$ for large distances, the inclusion of numerous angular momentum channels $J$ is necessary.}
	\label{fig:chargeintegraln1nc31}
\end{figure}

In this section, we study the screening of the impurity charge. It is convenient to examine the accumulated induced charge from  the Landau impurity  band $N$ within distance $d$ from the origin, defined as $Q_N (d) \equiv 2 \pi \int_{0}^{d} \rho_N (r) r dr$. Subtracting it from the total charge within the same distance in the absence of impurity, we obtain a charge difference that measures how much the impurity affects charge profiles within distance $d$:
\begin{equation}
	\Delta Q_N (d) = 2 \pi \int_{0}^{d} \left( \rho_N (r) - \frac{1}{2 \pi \ell^2} \right) r dr.
\end{equation}
For a large distance $d_s$, the influence of  the impurity vanishes, $\Delta Q_N(d_s) \approx 0$. Hence, $d_s$ may be interpreted as the screening length. As coupling strength $g$ increases, screening length $d_s$ is expected to increase.   A numerical result of the  charge difference $\Delta Q_N (d)$ for the impurity band $N=1$, plotted in Fig. \ref{fig:chargeintegraln1nc31}, supports this expectation.  Also, we observe that there is no sudden change in $\Delta Q_N (d)$ near the critical coupling strength $g_c = 1/2$, similar to the peak behavior in the induced density.

\section{Impurity cyclotron resonance}
\label{section:impurity_cyclotron}

 Impurity cyclotron resonance \cite{Goldman1986} may be used  to detect the discrete  energy levels in the energy spectrum. The optical matrix elements  between  the {\it graphene Landau level  states} in the absence of impurity ($g=0$) are evaluated by using the formula $\vec{j} = v_F \vec{\sigma}$ \cite{Beenakker2008}:
\begin{eqnarray}	
	\langle \psi_{n,m} |   \sigma_x |  \psi_{n',m'}  \rangle = M_{nn'}\delta_{mm'}.
\end{eqnarray}
Here, we consider the current along the $x$ axis.   (We recall that graphene is in the $xy$ plane.) The explicit matrix elements for optical selection rules are given in Tables \ref{table_optical_x}, which implies that $M_{n n'}$ is  non zero  only for $\Delta n = n' - n = \pm 1$. Combined with the implied rule of the Kronecker delta $\delta_{mm'}$, we can infer that the allowed transitions must satisfy either $\Delta J = 1$ or $\Delta J = -1$.

However, in the presence of a Coulomb potential, the optical matrix elements  must be  evaluated using {\it Landau impurity band states}. We find
\begin{eqnarray} 
&&\mathbf{T}_{(N, J) \rightarrow (N', J')} =\langle \Psi_{N, J} |   \sigma_x  |  \Psi_{N', J'}  \rangle \nonumber\\
&=& \sum_{n, n'}  (C_n^{N,J})^*  \langle \psi'_{n,J} \vert \sigma_x \vert \psi'_{n',J'} \rangle C_{n'}^{N',J'} \nonumber\\  
&=& \sum_{n, n'}  (C_n^{N,J})^* M_{n, n'} \delta_{|n|-J,|n'|-J'} C_{n'}^{N',J'} \nonumber\\   
&=& \sum_{n, n'}  (C_n^{N,J})^* \left[ M^{ (+1)}_{n, n'} \delta_{J+1,J'} + M^{ (-1)}_{n, n'} \delta_{J-1,J'}\right] C_{n'}^{N',J'},\nonumber\\
\label{impurity_T}		
\end{eqnarray}
where 
$M^{ (\pm 1)}_{n, n'} = \pm i c_n c_{n'} \delta_{|n| \mp 1, |n'|} \text{sgn}(n)$ 
correspond to angular momentum    $J$  increasing or decreasing by $1$.


\begin{table}[H]
	\setlength{\tabcolsep}{0.4em}
	{\renewcommand{\arraystretch}{2}
		\begin{tabular}{|c|ccccccccc}
			\hline
			\diagbox[width=\dimexpr \textwidth/16+2\tabcolsep\relax, height=1cm]{ $n$ }{$n'$}
			& $\ldots$ & $-3$ & $-2$ & $-1$ & 0 & 1 & 2 & 3 & $\ldots$\\
			\hline
			$-2$ & $\ldots$ & $\frac{i}{2}$   & 0 & $-\frac{i}{2}$ & 0 &$-\frac{i}{2}$ & 0 &  $-\frac{i}{2}$ & $\ldots$\\
			$ - 1$ & $\ldots$   & 0 &  $\frac{i}{2}$  & 0 & -$\frac{i}{\sqrt{2}}$ & 0 & $-\frac{i}{2}$  & 0 & $\ldots$\\
			$0 $ & $\ldots$ &   0 &  0 & $\frac{i}{\sqrt{2}}$  & 0 & -$\frac{i}{\sqrt{2}}$ & 0  & 0  & $\ldots$\\
			$ 1$ & $\ldots$   & 0 &  $\frac{i}{2}$  & 0 & $\frac{i}{\sqrt{2}}$ & 0 & $-\frac{i}{2}$  & 0 & $\ldots$\\
			$2$& $\ldots$ & $\frac{i}{2}$   & 0 & $\frac{i}{2}$ & 0 &$\frac{i}{2}$ & 0 &  $-\frac{i}{2}$ & $\ldots$\\
			\hline
	\end{tabular}}
	\caption{The matrix elements $M_{nn'}$ between the basis states, which correspond to the $\sigma_x$ optical transition, are displayed.
		\label{table_optical_x}}
\end{table}

\begin{figure}[H]
	\centering
	\includegraphics[width=\linewidth]{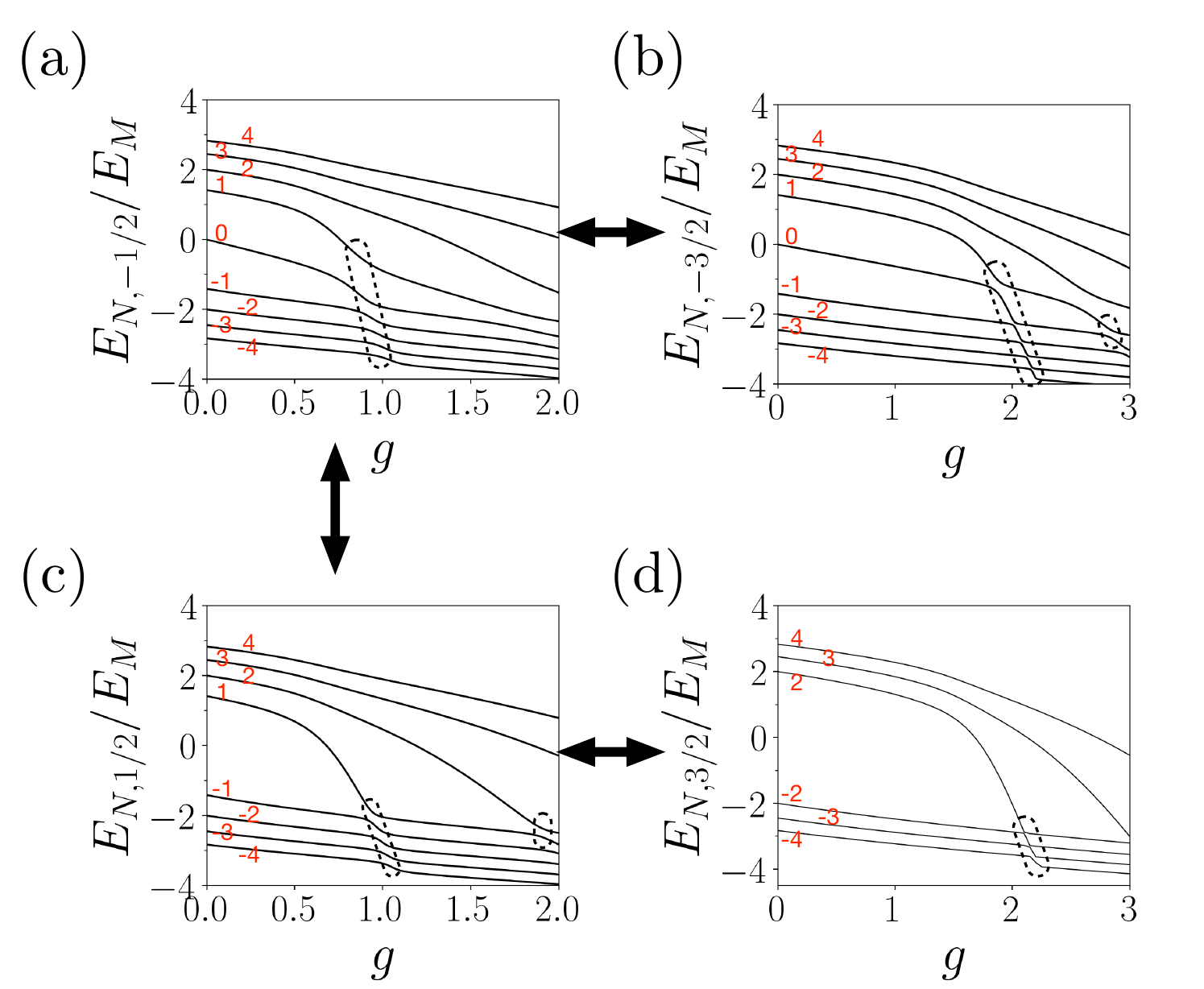}
	\caption{Energy spectra of (a) $J = -1/2$, (b) $J = -3/2$, (c) $J = 1/2$, and (d) $J = 3/2$ are plotted with $N_c = 2501$.   Multiple anticrossings occur in the dashed oval regions.  Each number attached to a curve indicates the corresponding  value of  $N$.    Thick arrows between the plots indicate that optical transitions ($\Delta J=\pm 1$) between the impurity band states shown in each plot are possible. }
	\label{new_anticrossing}
\end{figure}

The form $\mathbf{T}_{(N, J) \rightarrow (N', J')}$ above suggests the possibility of anomalous transitions within the same impurity band, i.e.,  $\Delta N = 0$.  In the limit $g=0$ the condition $\Delta N = 0$ changes into $\Delta n=0$, which is not optically  allowed, as mentioned above.   However, $\Delta N = 0$ is possible  at numerous finite values of $g$ because  two impurity Landau levels may cross each other \cite{Sobol2016}, as shown in Fig. \ref{new_anticrossing}.   We compared our numerical results with the eigenspectrum obtained using the shooting method to solve the Dirac equation as described in Ref. \cite{Sobol2016}, and obtained similar results using  $N_c=201$.

\begin{figure}[H]
   \centering
   \includegraphics[width=0.8\linewidth]{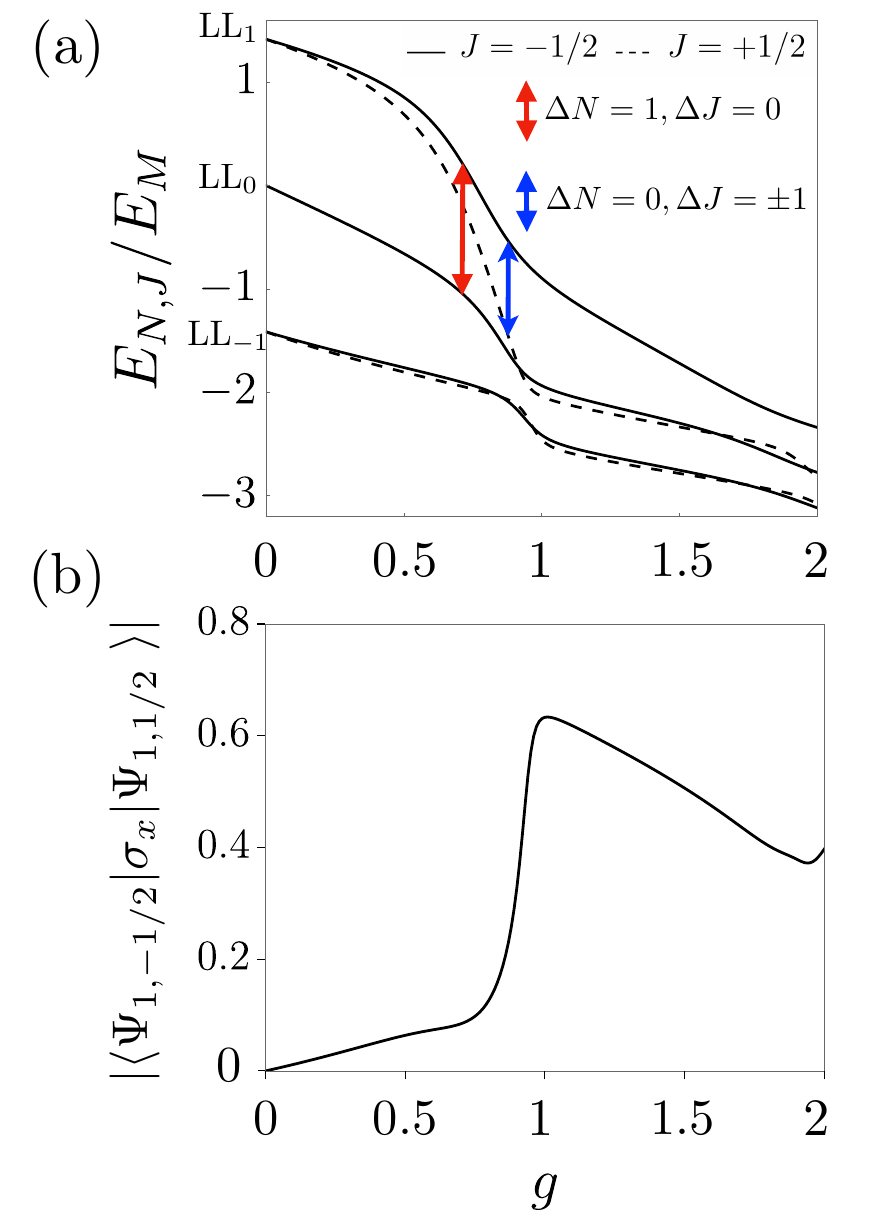}
   \caption{(a)  Illustration of forbidden   (red arrow with $\Delta J=0$) and anomalous (blue arrow with $\Delta J=\pm 1$)  optical transitions. (b) Optical matrix element $\big\vert \langle \Psi_{1, -1/2} \vert \sigma_x \vert \Psi_{1, 1/2} \rangle \big\vert  $ as a function of $g$.
  	Both of these plots are 
   	 computed with $N_c=2501$.}
    \label{Exp}
\end{figure}

Let us analyze an example of an anomalous optical transition to gain a better understanding. Its matrix element is given by
\begin{equation}
	\mathbf{T}_{(1, -1/2) \rightarrow (1, 1/2)} =  \langle \Psi_{1, -1/2} \vert \sigma_x \vert \Psi_{1, 1/2} \rangle.
\end{equation}
This transition is depicted in  Fig. \ref{Exp}(a), and 
the dependence of $\vert \mathbf{T}_{(1, -1/2) \rightarrow (1, 1/2)}  \vert$ on $g$ is plotted in Fig. \ref{Exp}(b).  
We observe the following properties. 
(i) For small values of coupling strength $g$, this optical matrix element is small. It can be explained by noting that with a small coupling strength, this optical matrix element is approximated by transition between graphene  Landau level states,  $\vert\psi'_{nJ}\rangle=\vert \psi'_{1, 1/2} \rangle\rightarrow \vert \psi'_{1, -1/2}  \rangle$, which is forbidden because $\Delta n=0$.
(ii) For a strong coupling strength, such as with $g > 1.5$, the transition  $\vert \psi'_{-1, 1/2} \rangle\rightarrow \vert \psi'_{0, -1/2}  \rangle$, which is allowed because $\Delta n= -1$, contributes significantly to $\mathbf{T}_{(1, -1/2) \rightarrow (1, 1/2)}$.
(iii) There is a crossover in $\mathbf{T}_{(1, -1/2) \rightarrow (1, 1/2)}$ as a function of $g$, that occurs around $g = 0.8$, which is a consequence of strong Landau level mixing.

\section{Discussion and Conclusions}
\label{section:conclusion}

 We investigated  the induced density of the supercritical Coulomb impurity in the regime where filled  Landau impurity bands do not overlap
	 and  the effect of electron-electron interactions is significantly reduced. 
The strong coupling between  graphene Landau level states by the impurity potential is non trivial and can lead to several anomalous  effects.   The induced density of a filled Landau impurity band can exhibit a sharp peak near the impurity center, much narrower than the magnetic length.  However,  due to strong coupling between graphene  Landau levels,  this  
 peak can be located away from the center of the impurity, depending on  the properties of different  Landau impurity  bands.
We found, like in the presence of discrete scale invariance, that states with angular momentum $J \leq g$ strongly contribute to the induced density near $r=0$.  In addition, the impurity charge is screened despite  the Landau impurity band being completely filled.
We also showed that additional impurity cyclotron resonances exist that involve the anticrossing of Landau impurity band states.

While it is desirable to conduct a Hartree-Fock calculation~\cite{MacDonald1993}, we do not anticipate qualitative changes in the induced density of a filled Landau impurity band, although some quantitative adjustments in the results are expected~\cite{Kim2014Impurity}.
A scanning tunneling microscope~\cite{Andrei2012} may be useful in investigating the anomalous induced density. Impurity cyclotron measurements~\cite{Goldman1986} may also prove useful.

\appendix
\section{Discrete scale invariance}\label{App1}

The following  simple example illustrates what  discrete scale invariance  is.  
Consider the function 
\begin{equation}
	f(x)=x^{\nu}
\end{equation}
with  an {\it imaginary} scaling exponent $\nu=i\eta$.  This function  displays discrete scale invariance involving the exponent $\nu$ as follows:

\begin{equation}
	x\rightarrow \lambda x, \quad  \lambda=e^{\pm i\frac{\pi}{\nu}}=e^{\pm \frac{\pi}{\eta}}.
\end{equation}
We can rewrite the function as
\begin{equation}
	x^{\nu}=e^{i \eta  \text{ln} x}= \cos(\eta  \text{ln}x)+i \sin(\eta  \text{ln}x).
	\label{Log}
\end{equation}
This function exhibits {\it log-periodic} oscillations as a function of $x$.
In graphene discrete scale invariance also shows up~\cite{Gamayun2009} in the complex eigenenergies for $g>|J|$, 
\begin{eqnarray}
	E_n= e^{\frac{-\pi}{\eta}} E_{n-1},
\end{eqnarray} 
where $J$ is the half-integer angular momentum quantum number.   One of the key factors of this mathematical structure is the appearance of the same exponent as in the log-periodicity  of the wavefunctions given by Eq. (\ref{Log}), with the exponent:
\begin{eqnarray}
	\eta=\sqrt{g^2-J^2}.
\end{eqnarray} 
For $g>|J|$, the exponent $\eta$ is greater than zero, and the wavefunctions display log-periodic oscillations. The larger the coupling constant is, the more angular momentum channels are affected.

\section{Eigenstates of an Ordinary Two-Dimensional Electron gas in Magnetic Fields}\label{Twostates}

In polar coordinates, the two-dimensional Landau level wave functions of an ordinary two-dimensional electron gas \cite{Yoshioka} are given for $n\ge 0$ and $m\ge 0$ by 
\begin{equation}
	\phi_{n,m} (\vec{r}) = A_{n,m} \exp \left[ i (\vert n \vert - m) \theta - \frac{r^2}{4 \ell^2} \right] \left( \frac{r}{\ell} \right)^{\alpha} L^{\alpha}_{\beta} \left( \frac{r^2}{2 \ell^2}\right),
\end{equation}
where $L^\alpha_\beta(x)$ are generalized Laguerre polynomials.
 The $z$ component of the angular momentum of $\phi_{n,m} (\vec{r})$ is $L_z=-i\partial/\partial \theta=\hbar(n-m)$. 
Note that, in contrast to graphene states, these states are one-component wave functions.  Also, the definition of the $z$ component of angular momentum is different.
Using the orthogonality of Laguerre polynomials
\begin{equation}
	\int_{0}^{\infty} t^\alpha  e^{-t} L^\alpha_m (t) L^\alpha_n (t) = \frac{\Gamma \left(n+\alpha+1\right)}{n!} \delta_{m,n},
	\label{orthogonality}
\end{equation}
the normalization factor is derived as
\begin{equation}
	\begin{aligned}
		A_{n,m} &= \frac{1}{\ell} \left({2\pi \; 2^\alpha \frac{\Gamma ({\beta + \alpha +1}) }{\beta!}} \right) ^{-1/2},
	\end{aligned}
	\label{Anm}
\end{equation}
with $\alpha = \vert m - \vert n \vert \vert$ and $\beta = \left(m + \vert n \vert - \alpha\right)/2$. 
All the  states $	\phi_{n,m} (\vec{r}) $ decay exponentially as $e^{-r^2/2\ell^2}$.
One can show the following identity  for the expectation value of $r^2$:
\begin{equation}
	\langle \phi_{n,m} |r^2 | \phi_{n,m}   \rangle =2 \ell^2(n+m+1).
	\label{posit}
\end{equation}

\section{Eigenstates}\label{states}

\begin{figure}[H]
	\centering
	\includegraphics[width=0.7\linewidth]{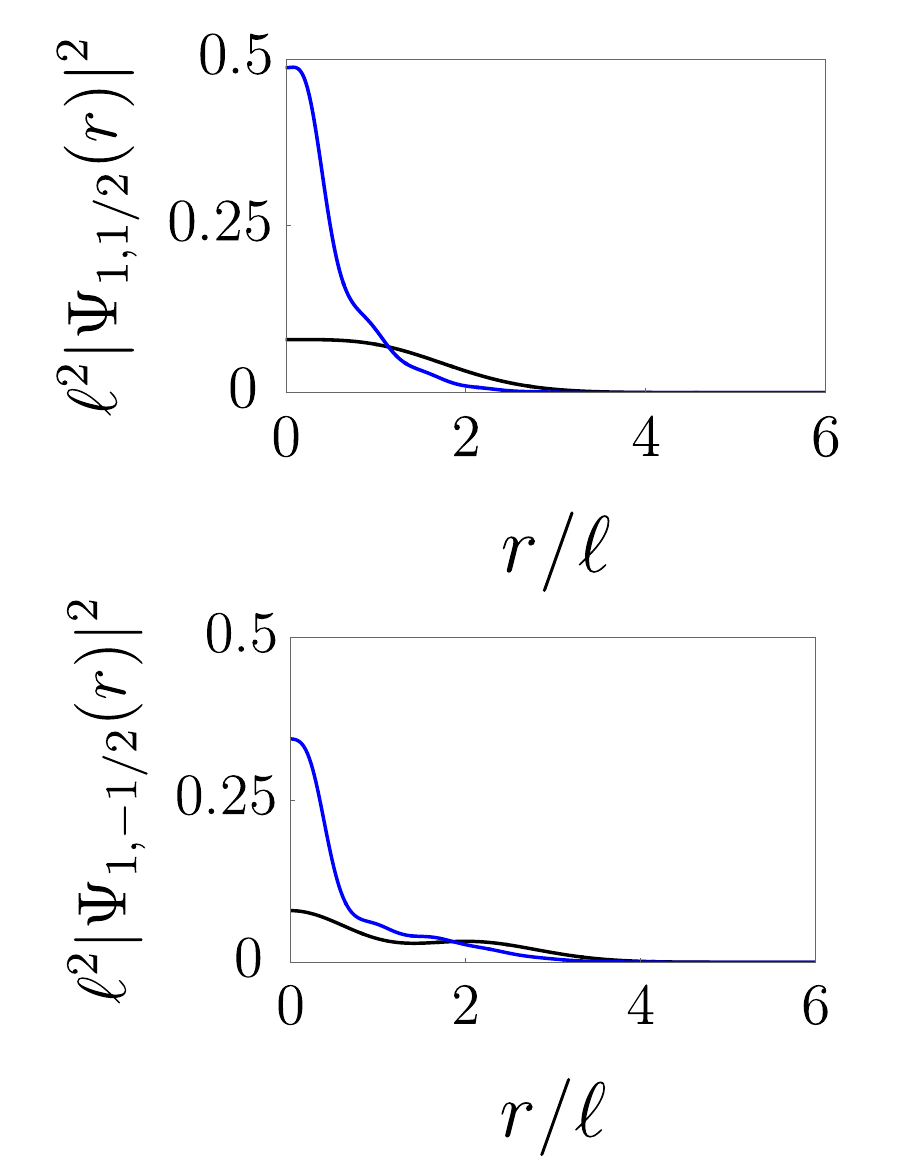}
	\caption{Probability density of the $\Psi_{1,1/2}(r)$ and   $\Psi_{1,-1/2}(r)$ states for $g=0$ (black line) and $0.55$ (blue line).  Other states in the Landau impurity band $N=1$ undergo minimal changes.}
	\label{n=1}
\end{figure}

\begin{figure}[H]
	\centering
	\includegraphics[width=0.55\linewidth]{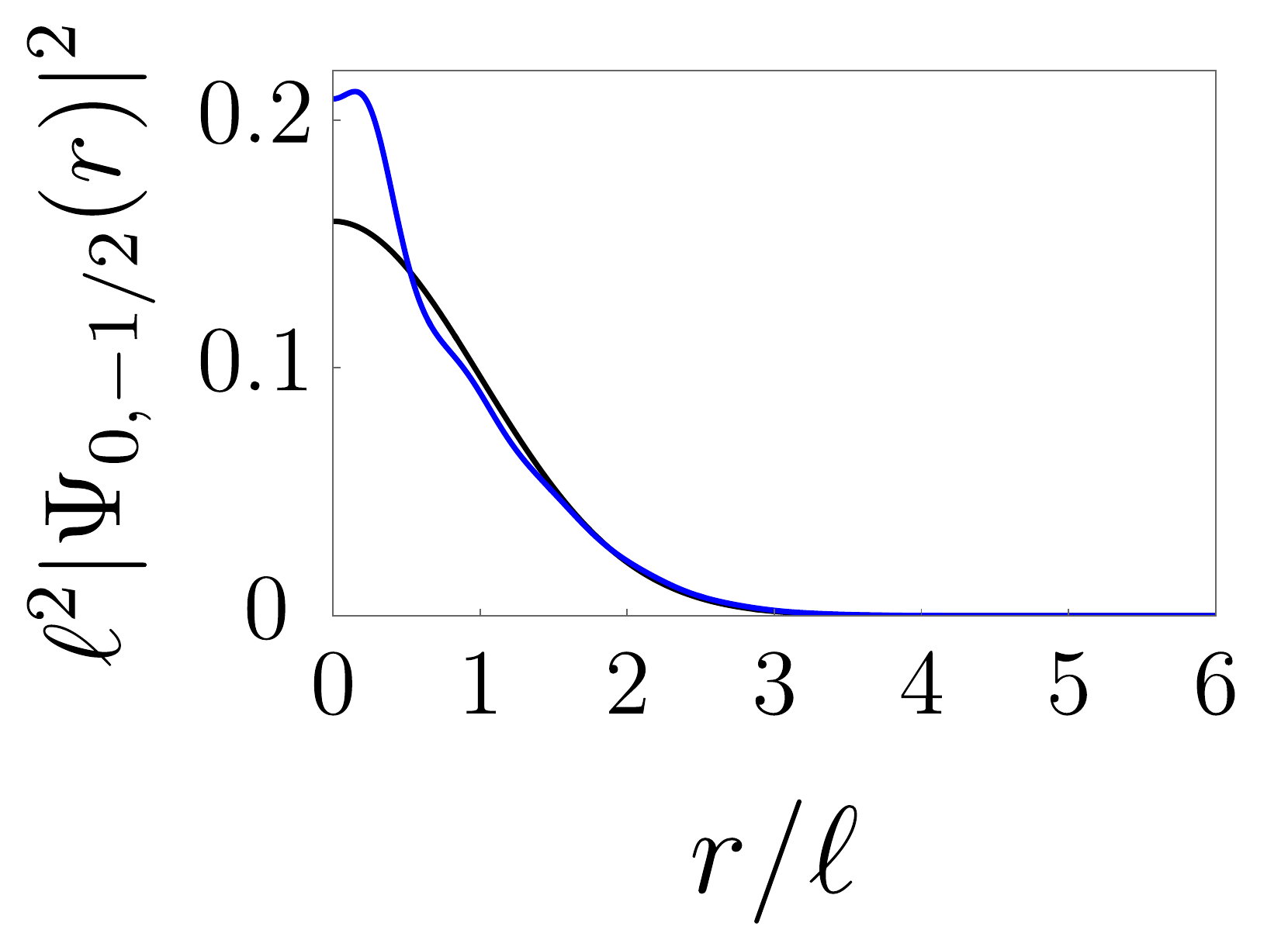}
	\caption{Probability density $|\Psi_{0,-1/2}(r)|^2$ for $g=0$ (black line) and $0.55$ (blue line). Only this state is significantly affected by the Coulomb field, while other states in the Landau impurity band $N=0$ undergo minimal changes.}
	\label{n=0}
\end{figure}

\begin{figure}[H]
	\centering
	\includegraphics[width=0.99\linewidth]{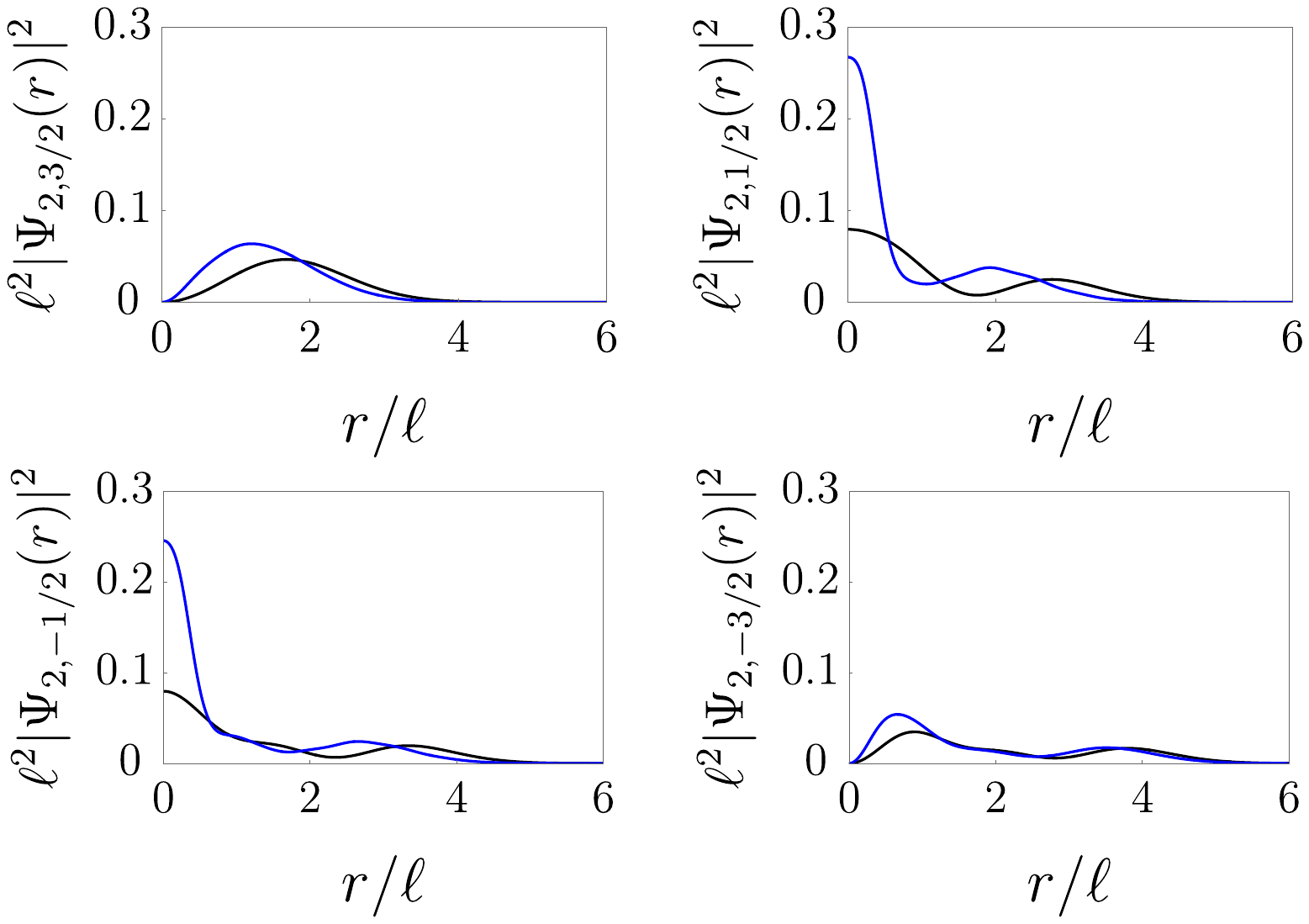}
	\caption{Probability densities of $\Psi_{2,3/2}(r)$, $\Psi_{2,1/2}(r)$, $\Psi_{2,-1/2}(r)$, and  $\Psi_{2,-3/2}(r)$ for $g=0$ (black line) and $0.55$ (blue line). Other states in the Landau impurity band $N=2$ barely change.}
	\label{n=2}
\end{figure}

\begin{figure}[h!]
	\centering	\includegraphics[width=\linewidth]{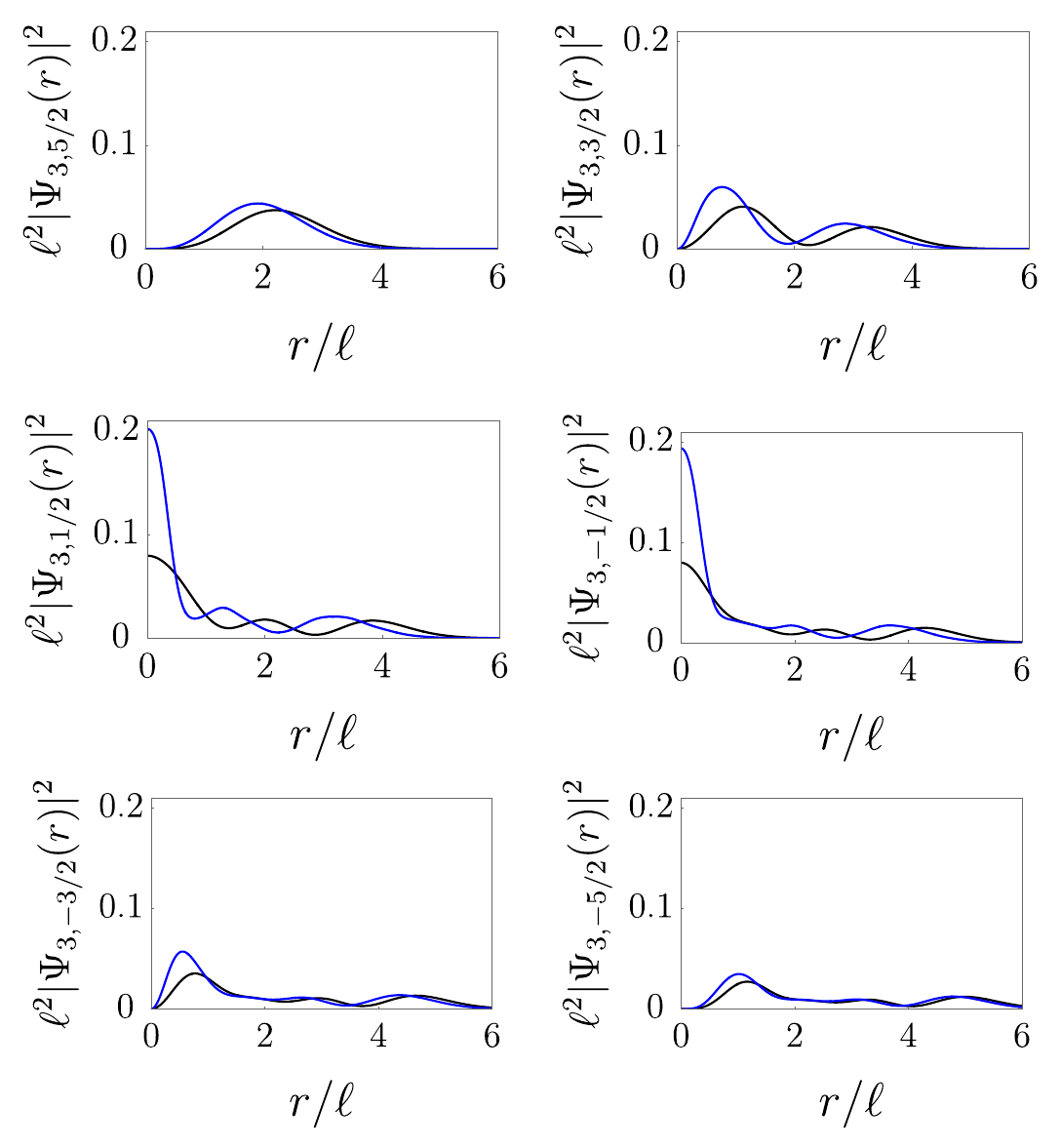}
	\caption{Probability densities of $\Psi_{3,5/2}(r)$, $\Psi_{3,3/2}(r)$, $\Psi_{3,1/2}(r)$,  $\Psi_{3,-1/2}(r)$, $\Psi_{3,-3/2}(r)$, and   $\Psi_{3,-5/2}(r)$ for $g=0$ (black line) and $0.55$ (blue line). Other states in the Landau impurity band $N=3$ barely change.}
	\label{n=3}
\end{figure}

We analyze the properties of different impurity band states. The following points are worth noting:
	\begin{enumerate}
	\item Probability distributions $|\Psi_{N,\pm1/2}(r)|^2$ for $J=\pm1/2$ are peaked at $r=0$ (see Fig. \ref{n=1}). In some cases, they are peaked away from $r=0$, as shown in Fig. \ref{n=0}. As a function of $g$, the wavefunctions do not display a sharp transition at $g_c=1/2$, unlike the zero magnetic field case.
	
	\item Probability distributions $|\Psi_{N,J}(r)|^2$ for $J\neq \pm1/2$ are peaked away from $r=0$ (see Figs. \ref{n=2} and \ref{n=3}). However, $\Psi_{N,J}(r)=0$ at $r=0$. Only states with $J=\pm 1/2$ are non zero at $r=0$.
\end{enumerate}

\pagebreak

\bibliography{ref_supercriticality.bib}

\end{document}